\documentclass{article}
\usepackage{arxiv}

\usepackage[utf8]{inputenc} 
\usepackage[T1]{fontenc}    
\usepackage{hyperref}       
\usepackage{url}            
\usepackage{booktabs}       
\usepackage{amsfonts}       
\usepackage{nicefrac}       
\usepackage{microtype}      
\usepackage{graphicx}
\usepackage{doi}

\usepackage{textcomp}

\usepackage{tikz}
\usepackage{booktabs}
\usepackage{optidef}
\usepackage{bm}

\usepackage{graphicx}
\usepackage{amsmath}
\usepackage[version=4]{mhchem}
\usepackage{siunitx}
\usepackage{longtable,tabularx}
\usepackage{glossaries}
\newacronym{SSTL}{SSTL}{Surrey Satellite Technology Ltd}

\newacronym{GEO}{GEO}{geostationary orbit}
\newacronym{LEO}{LEO}{low-Earth orbit}
\newacronym{GTO}{GTO}{geostationary transfer orbit}
\newacronym{LPO}{LPO}{lunar polar orbit}
\newacronym{SSO}{SSO}{Sun-synchronous orbit}
\newacronym{NRHO}{NRHO}{near-rectilinear halo orbit}

\newacronym{CC}{CC}{cone-clock}

\newacronym{COCP}{COCP}{continuous optimal control problem}
\newacronym{TPBVP}{TPBVP}{two-point boundary value problem}
\newacronym{MPBVP}{MPBVP}{multi-point boundary value problem}
\newacronym{CR3BP}{CR3BP}{circular restricted 3-body problem}
\newacronym{COV}{COV}{calculus of variations}
\newacronym{GVE}{GVEs}{Gauss variational equations}
\newacronym{PMP}{PMP}{Pontryagin's minimum principle}
\newacronym{NLP}{NLP}{nonlinear programming}
\newacronym{KKT}{KKT}{Karush-Kuhn-Tucker}

\newacronym{EP}{EP}{electric propulsion}
\newacronym{SB}{SB}{shape-based}
\newacronym{NC}{NC}{neurocontroller}
\newacronym{CLFD}{CLFD}{closed-loop feedback-driven}
\newacronym{ZEM/ZEV}{ZEM/ZEV}{zero-effort-miss/zero-effort-velocity}
\newacronym{DAG}{DAG}{directional adaptive guidance}

\newacronym{GA}{GA}{genetic algorithm}
\newacronym{PSO}{PSO}{particle swarm optimisation}
\newacronym{ACO}{ACO}{ant colony optimisation}
\newacronym{ICEA}{ICEA}{improved cooperative evolutionary algorithm}

\newacronym{SBT}{SBT}{stochastic batch training}
\newacronym{DT}{DT}{deterministic training}
\newacronym{OD}{OD}{orbit determination}
\newacronym{OI}{OI}{orbit insertion}
\newacronym{EX}{EX}{execution}

\newacronym{LVLH}{LVLH}{local-vertical-local-horizontal}
\newacronym{RTN}{RTN}{radial-transverse-normal}
\newacronym{ECI}{ECI}{Earth-centred inertial}
\newacronym{MCI}{MCI}{Moon-centred inertial}
\newacronym{SOI}{SOI}{sphere of influence}

\newacronym{RL}{RL}{reinforcement learning}
\newacronym{ML}{ML}{machine learning}
\newacronym{NN}{NN}{neural network}
\newacronym{MDP}{MDP}{Markov decision process}
\newacronym{AC}{AC}{actor-critic}
\newacronym{ELM}{ELM}{extreme learning machine}
\newacronym{SLFN}{SLFN}{single-layer feed-forward network}
\newacronym{DDPG}{DDPG}{deep deterministic policy gradient}
\newacronym{PPO}{PPO}{proximal policy optimisation}
\newacronym{TRPO}{TRPO}{trust-region policy optimisation}
\newacronym{ReLU}{ReLU}{rectified linear unit}
\newacronym{RBF}{RBF}{radial basis function}
 
\newacronym{RAAN}{RAAN}{right ascension of the ascending node}
\newacronym{MEE}{MEE}{modified equinoctial element}
\newacronym{COE}{COE}{classical orbital element}
\newacronym{ROE}{ROE}{relative orbital element}

\newacronym{SRP}{SRP}{solar radiation pressure}

\newacronym{VMMO}{VMMO}{Volatile Mineralogy Mapping Orbiter}
\newacronym{ADCS}{ADCS}{Attitude Determination and Control System}
\newacronym{GNC}{GNC}{guidance, navigation and control}
\newacronym{MC}{MC}{Monte Carlo}

\newacronym{RLC}{RLC}{Reinforced Lyapunov Controller}
\newacronym{DACX}{DACX}{differential algebra based successive convex optimisation}
\newacronym{SCVX}{SCVX}{successive convex optimisation}
\newacronym{SCP}{SCP}{Sequential Convex Programming}
\newacronym{FOH}{FOH}{first-order-hold discretisation}
\newacronym{LLO}{LLO}{low-altitude lunar orbit}
\newacronym{eLLO}{eLLO}{extremely-Low Lunar Orbit}
\newacronym{ZOH}{ZOH}{zero-order-hold}
\newacronym{LRO}{LRO}{Lunar Reconnaissance Orbiter}
\newacronym{ICRF}{ICRF}{international celestial reference frame}
\newacronym{DCM}{DCM}{direction cosine matrix}
\newacronym{LME2000}{LME2000}{lunar mean equator of date J2000}
\newacronym{GRAIL}{GRAIL}{Gravity Recovery And Interior Laboratory}

\newacronym{SDF}{SDF}{Signed Distance Field}
\newacronym{AD}{AD}{Automatic Differentiation}

\usepackage{algorithm}
\usepackage[noend]{algpseudocode}

\title{Using the Translation Theorem for the Automated Stationkeeping of Extremely-Low Lunar Missions}

\date{}

\usepackage{authblk}

\author[1]{%
	Jack Yarndley\thanks{\texttt{Corresponding Author: jyar540@aucklanduni.ac.nz}}\hspace{1.2mm}%
}
\author[2]{%
	{Martin Lara}%
}
\author[3]{%
	{Harry Holt}%
}
\author[1]{%
	{Roberto Armellin}%
}

\affil[1]{Te P\=unaha \=Atea -- Space Institute, University of Auckland, Auckland 1010, New Zealand}
\affil[2]{Scientific Computation Research Center, University of La Rioja, Logroño, 93 26006, Spain}
\affil[3]{Advanced Concepts Team, ESTEC, European Space Agency, Noordwijk, 2201 AZ, The Netherlands}

\renewcommand{\headeright}{}
\renewcommand{\undertitle}{Preprint}
\renewcommand{\shorttitle}{\textit{Using the Translation Theorem for the Automated Stationkeeping of Extremely-Low Lunar Missions}}

\hypersetup{
pdftitle={Using the Translation Theorem for the Automated Stationkeeping of Extremely-Low Lunar Missions},
pdfsubject={physics.space-ph},
pdfauthor={Jack Yarndley, Martin Lara, Harry Holt, Roberto Armellin},
pdfkeywords={translation theorem, dynamical averaging, space, lunar missions, convex programming, mission design},
}

\usepackage[numbers]{natbib}

\begin{document}

\maketitle
\begin{abstract}
\Glspl{eLLO} (altitudes $\leq 50$ km) exhibit severe perturbations due to the highly non-spherical lunar gravitational field, presenting unique challenges to orbit maintenance. These altitudes are too low for the existence of stable `frozen' orbits, and naive stationkeeping methods, such as circularization, perform poorly. However, mission designers have noticed a particular characteristic of low lunar orbits, which they have found useful for stationkeeping and dubbed the ``translation theorem'', wherein the eccentricity vector follows a predictable monthly pattern that is independent of its starting value. We demonstrate this feature results from the low orbital eccentricity combined with the dominant effect of a particular subset of sectoral and tesseral harmonics. Subsequently, automated stationkeeping strategies for \glspl{eLLO} are presented, utilizing this theorem for eccentricity vector control. Several constraints within the eccentricity vector plane are explored, including circular, annular, and elevation-model derived regions, each forming distinct stationkeeping strategies for varying orbital configurations. Subsequently, the optimal control profiles for these maneuvers within the eccentricity plane are obtained using \gls{SCP}. The proposed strategies offer computational simplicity and clear advantages when compared to traditional methods and are comparable to full trajectory optimization.
\end{abstract}

\glsresetall

\section{\label{section_introduction}Introduction}
The design of long-duration lunar missions is inherently challenging, particularly due to the highly nonlinear lunar gravity field. This introduces significant perturbations to the Keplerian gravity model. Additionally, unlike in our terrestrial counterpart, the absence of a lunar atmosphere provides opportunities for spacecraft to explore \glspl{eLLO}, with altitudes below 50 km. These orbits present unique scientific and technological opportunities, such as high-resolution imaging, high-bandwidth communication, magnetometry, and gravitational studies. However, the maintenance and stationkeeping designs of \glspl{eLLO} are particularly intricate due to the increasing dominance of nonlinear gravitational effects at such altitudes, which render traditional orbital control strategies ineffective.

To simplify these challenges, the gravitational perturbations are often modeled using only the terms that drive long-term dynamics. Zonal harmonic models are widely used in this context for the design of non-resonant orbital configurations \citep{CuttingFrautnickBorn1978, RosboroughOcampo1991, shapiroPhasePlaneAnalysis1996, hootsHistoryAnalyticalOrbit2012}, as they capture the primary long-term effects of a nonlinear gravity model. This approach is supported by observing that the short-period terms have negligible amplitude, meaning they can be effectively removed through classical averaging methods \citep{Brouwer1959, Kozai1959}. This reduction process preserves the long-period zonal harmonics whilst eliminating most sectoral and tesseral contributions \citep{Kaniecki1979, Kaula1966}. Further simplifications can be achieved via double-averaging techniques, which exploit the rapid rotation of bodies like Earth to filter out the longitude-dependent tesseral and sectoral terms \citep{metrisSemianalyticalTheoryMean1995, deleflieLongPeriodVariationsEccentricity2006, Palacian2007, laraAveragingTesseralEffects2013, Lara2021}. However, these methods are much less effective for the Moon due to its significantly slower rotation rate.

In low lunar orbits, the amplitude of longitude-dependent terms—scaled by the ratio of mean motion to the Moon’s slow rotation rate—becomes significant and cannot be neglected \citep{Kaula1966, Garfinkel1965b, Lara2011}. This effect is amplified by the comparable magnitude of the zonal oblateness coefficient ($J_2$) to various sectoral and tesseral coefficients. Therefore, ignoring the amplitude of the longitude-dependent terms, which are removed in the second averaging process, may compromise the design of lower lunar orbits \citep{knezevicPerturbationTheoryLow1995, holtExtremelyLowaltitudeLunar2024}. This is even more important in the case of $m$-monthly (the lunar equivalent of $m$-daily) terms because they are the only ones having terms independent of eccentricity. The resulting sectoral and tesseral contributions lead to a recurring modulation in the argument of perilune and eccentricity, forming recognizable repeatable patterns \citep{holtExtremelyLowaltitudeLunar2024, foltaLunarFrozenOrbits2006, BeckmanLamb2007}. This phenomenon, first identified by \citep{BeckmanLamb2007}, was described as the ``translation theorem'' though it lacked a detailed explanation at the time. The full explanation of this phenomenon, which builds on preliminary work by the authors \citep{holtExtremelyLowaltitudeLunar2024} that considers circular stationkeeping regions, is the first major contribution of this work. 

The analytical justification of the translation theorem emerges from a linearization of the perturbation dynamics under the Moon’s gravitational field \citep{Kaula1966, Lara2011}. By considering the mean motion dynamics in combination with the $m$-monthly corrections, the evolution of the eccentricity vector in the orbital plane can be described analytically. These corrections are specifically associated with sectoral and tesseral harmonics of odd degree, revealing that the $m$-monthly terms neither depend on the argument of perilune ($\omega$) nor require high-order eccentricity terms for accurate prediction. As a result, the mean eccentricity vector undergoes a predictable, nearly repetitive modulation with a distinct monthly pattern largely independent of the initial argument of perilune. This regularity is the foundation of the translation theorem and provides valuable insight into low-altitude orbital evolution. In essence, the translation theorem connects the long-term evolution of mean orbital elements with the predictable short-term dynamics of the eccentricity vector. This relationship enables the approximation of the eccentricity vector trajectory, offering a foundation for developing stationkeeping strategies. 

In past studies, significant efforts have focused on identifying stable frozen or quasi-frozen lunar orbits to mitigate the effects of gravitational perturbations. Such orbits exhibit minimal long-term variation under the influence of the Moon's nonlinear gravity; however, they are generally impractical at altitudes below 50 km \citep{Konopliv1993, Lara2011, Russell2007, foltaLunarFrozenOrbits2006}. Many studies on stationkeeping strategies, such as those conducted for the \gls{LRO}, have predominantly addressed higher altitudes ($\ge$ $50$ km) \citep{BeckmanLamb2007, leonardiLowThrustNonlinearOrbit2024}, or relied on simplifications that may not yield feasible strategies. For instance, \citep{Singh2020} analyzed near-polar \glspl{eLLO} but permitted significant apolune growth in their approach, which is an issue for missions that intend for low average altitudes.

In terms of stationkeeping strategies at the lower altitudes ($< 50$ km) the \gls{GRAIL} mission presents if not the only case of \gls{eLLO} stationkeeping, particularly in the design of the extended mission \citep{sweetserDesignExtendedMission2012} and terminal phase of the mission \citep{wallaceLastDaysGRAIL2014}. The extended mission featured a mean altitude of $23.5$ km, and in the terminal phase, the periapsis altitude was permitted to drop to $\approx 2$ km above the lunar surface. The stationkeeping strategy utilized for this mission relied on graphical manipulation \citep{wallaceLowLunarOrbit2012} of the eccentricity vector to fall within prescribed eccentricity ranges for regularly spaced correction intervals between which single impulsive correction maneuvers were optimized.

Therefore, within this context, the proposal of an algorithm leveraging the translation theorem for the automated generation of fuel-optimal stationkeeping strategies for \gls{eLLO} is the second major contribution of this work. The utilization of the translation theorem allows for the design of optimized sequences of eccentricity vector translations for stationkeeping, which are subsequently converted into feasible control profiles for the spacecraft via \gls{SCP} \citep{chaiReviewOptimizationTechniques2019}. Overall, this approach achieves performance comparable to full trajectory optimization while maintaining computational efficiency so that it could be implemented onboard. The results presented in this work demonstrate the effectiveness of this method for various \glspl{eLLO}, providing a generalizable framework for the design and maintenance of low-altitude lunar orbits.

The paper is organized around the two major contributions introduced earlier. Section \ref{section_dynamical_formulation} presents the fundamental elements used throughout the manuscript, including the lunar dynamics and the characteristics of three \gls{eLLO} missions, which are used to demonstrate and justify the effectiveness of the proposed stationkeeping strategies. Next, in Section \ref{section_translation}, the translation theorem is introduced and justified. This section considers a classical doubly averaged dynamical model that captures the long-term behavior of the eccentricity vector, which is complemented by an analysis of the monthly contributions from the sectoral and tesseral harmonics to reveal the translation property. This theoretical foundation enables the development of efficient eccentricity vector control strategies, as detailed in Section \ref{section_stationkeeping}. Specifically, this includes the definition of mission-specific admissible regions for station keeping, referred to as eccentricity vector control regions; establishing the translation magnitudes required for station keeping; searching for initial conditions that minimize stationkeeping effort; and computing the actual stationkeeping maneuvers using \gls{SCP}. This methodology is demonstrated with application to three distinct mission scenarios: the Lunar Reconnaissance Orbiter \citep{BeckmanLamb2007}, the proposed Binar Prospector \citep{Buchan2022, Downey2022}, and the proposed `SER3NE' orbiter \citep{SER3NESELENESEXPLORER}.

\section{\label{section_dynamical_formulation}Lunar Environment}
In this section, we introduce the orbital dynamics relevant to \glspl{eLLO}, and introduce the appropriate reference frames and their associated coordinate systems. Then, three missions within the low-lunar environment are introduced as test cases, and subsequently, an appropriate truncation of spherical harmonics in the gravity model is made.

\subsection{\label{section_orbiter_dynamics}Dynamics and perturbations in low lunar orbits}
An accurate description of the orbital dynamics requires a well-defined inertial reference frame. To this end, the lunar mean equator of date J2000 (LME2000) frame is introduced, with its unit vectors ($c_1$, $c_2$, $c_3$) defined as follows: (i) $c_1$ acts as the $+x$ axis and points toward the intersection between the Moon's equator of date and the J2000 equator; (ii) $c_3$ acts as the $+z$ and points toward the Moon's north pole of date J2000; (iii) $c_2$ acts as the $+y$ axis and completes the right-hand frame. The origin of the LME2000 frame is located at the Moon's center of mass. 

However, as the Moon rotates around its axis, the non-spherical gravity field it produces does also. Therefore, a body-fixed frame rotating with the Moon must be used in order to model the gravity field. For analysis relating to the averaging process, and the translation theorem, a rotating frame aligned with the LME2000 frame at J2000 is used that rotates anti-clockwise around the $+z$ axis once every $27.3217$ days (the lunar sidereal month). However, this frame becomes inaccurate as the Moon's true rotation axis exhibits perturbations over time \citep{parkJPLPlanetaryLunar2021}. Therefore, depending on the format of the spherical harmonics provided, two very similar body-fixed ephemeris frames are used, the lunar PA421 \citep{folknerPlanetaryLunarEphemeris2009} and lunar PA440 \citep{parkJPLPlanetaryLunar2021} principal axis (PA) frames.

The spacecraft is subsequently modelled as a point mass. When considering dynamics in \glspl{eLLO}, there are a range of relevant perturbations to the Keplerian gravity model, including the non-spherical gravity field, third body effects (such as those from the Earth and Sun), and also solar radiation pressure. In this work, given the low altitudes of \glspl{eLLO}, the perturbations arising from the lunar non-spherical gravity field become dominant, so we will only consider its effects. In this model, the gravitational field is given by the potential
\begin{equation}\label{equation_gravitational_potential}
\begin{aligned}
    V = -\frac{\mu}{r} - R_{m},
\end{aligned}
\end{equation}
where $\mu$ is the gravitational parameter of the central body, $r$ the distance of the satellite to the origin, and $R_m$ are the perturbations due to the non-spherical gravity of the central body; in this case, the Moon. These perturbations can be written in the form of spherical harmonics as
\begin{equation}\label{equation_gravitational_harmonics}
\begin{aligned}
    R_{m} = \frac{\mu}{r}\sum_{j\geq2}\frac{\alpha^j}{r^j} \sum_{k=0}^{j} \left(C_{j,k}\cos k\lambda + S_{j,k}\sin k\lambda \right) P_{j,k}(\sin\varphi),
\end{aligned}
\end{equation}
where $\alpha$ is the reference radius of the central body over which the spherical harmonics are calculated, $\varphi$ and $\lambda$ the current latitude and longitude, $C_{j,k}$ and $S_{j,k}$ are the spherical harmonic coefficients, and $P_{i,j}$ are the associated Legendre polynomials of degree $i$ and order $j$. 

Data for the harmonic coefficients, and thus the model of the nonlinear lunar gravity field, is obtained from the GL0660B model \citep{konoplivJPLLunarGravity2013}. This model is obtained from the primary period of the \gls{GRAIL} mission and features a degree and order of up to 660 and is defined with respect to the PA421 frame.

\subsection{\label{section_mission_explainer}Missions within the low-lunar environment}
A range of different test cases relating to the stationkeeping of flown and proposed missions within the \gls{eLLO} environment are considered within this work. 
\begin{itemize}
    \item `LRO': The nominal mission flown by the Lunar Reconnaissance Orbiter \citep{BeckmanLamb2007}, maintaining a near-circular, polar orbit for a duration of 1 year is desired.
    \item `BINAR': The Binar Prospector mission, a proposed small satellite geophysical survey mission to identify lunar resources from orbit \citep{Buchan2022, Downey2022}. The resolution of the data depends on the altitude above the surface, so this should be as low as possible. The nominal duration of the mission is 90 days.
    \item `SER3NE': A proposed small satellite mission to perform gamma-ray and neutron spectroscopy \citep{SER3NESELENESEXPLORER}. Currently, an eccentric polar orbit is proposed. For lunar coverage, it is intended for the argument of perilune to naturally drift over the mission lifetime of 1 year. Limited information is publicly available on the exact desired orbital behavior, so some of the presented information is best guess based on what might be reasonable. 
\end{itemize}

\begin{table}[ht]
    \centering
    \begin{tabular}{lccc}
        \toprule
        & \textbf{LRO} \citep{BeckmanLamb2007} & \textbf{BINAR} \citep{Buchan2022, Downey2022} & \textbf{SER3NE} \citep{SER3NESELENESEXPLORER} \\
        \midrule
        \multicolumn{4}{l}{\textit{Nominal Parameters}} \\
        \quad Altitude & $50 \times 50$ \si{km} & $18 \times 18$ \si{km} & $20 \times 170$ \si{km} \\
        \quad Eccentricity & $0.0$ & $0.0$ & $0.0409$ \\
        \quad Inclination & $\ang{90}$ & $\ang{90}$ & $\ang{90}$ \\
        \quad Argument of Perilune & --- & --- & $\ang{-90}$ \\
        \quad Mission duration & 365 days & 90 days & 365 days \\
        \multicolumn{4}{l}{\textit{Parameter Ranges}} \\
        \quad Perilune altitude & $[30\,\si{km}, 50\,\si{km}]$ & $[\text{no impact}, 18\,\si{km}]$ & $[12\,\si{km}, 28\,\si{km}]$ \\
        \quad Eccentricity & $[0.0, 0.0112]$ & $[0.0, \text{no impact}]$ & $[0.0355, 0.0453]$ \\
        \quad Inclination & $[\ang{80}, \ang{100}]$ & $[\ang{70}, \ang{110}]$ & $[\ang{80}, \ang{100}]$ \\
        \quad Argument of Perilune & --- & --- & $[\ang{-135}, \ang{-45}]$ \\
        \bottomrule
    \end{tabular}
    \caption{Orbital parameters for cases used within this work.}
    \label{table_test_cases}
\end{table}

The desired orbit parameters and limits for these missions are presented in Table~\ref{table_test_cases}. Each mission has a nominal orbit and additional ranges in which the orbital parameters can lie. The parameter ranges work to define the region of orbital phase space in which each mission can exist. For the BINAR mission, slightly modified perilune and eccentricity ranges are permitted; rather than a fixed parameter range, the condition that the orbit does not impact the lunar surface is instead applied. For the purposes of this investigation, the `no impact' condition is interpreted such that the mission remains more than $1$ km above the lunar surface.

Although all missions are in different stages, for this study a nominal start time of 2026-01-01T00:00:00.0 UTC is chosen for all mission configurations. Since the dynamical model solely considers the lunar spherical harmonic model (not including other time-dependent perturbations such as solar radiation pressure or the effects of other bodies), the start time only determines the rotation between the LME2000 and PA440 frames. Therefore, a mapping can be directly constructed from these results to other desired starting times.

\subsection{\label{section_gravity_truncation}Truncation of the gravity model}
    An appropriate truncation for the lunar gravity model must be selected for our analysis. Too high of an order (for example, the full GL0660B model) and the gravity model can be extremely expensive to evaluate, making a comprehensive search of orbital configurations impractical. Conversely, too low of an order will fail to accurately capture the dynamical behavior of \glspl{eLLO}. 

\begin{figure}[!ht]
\centering
\includegraphics[width=\textwidth]{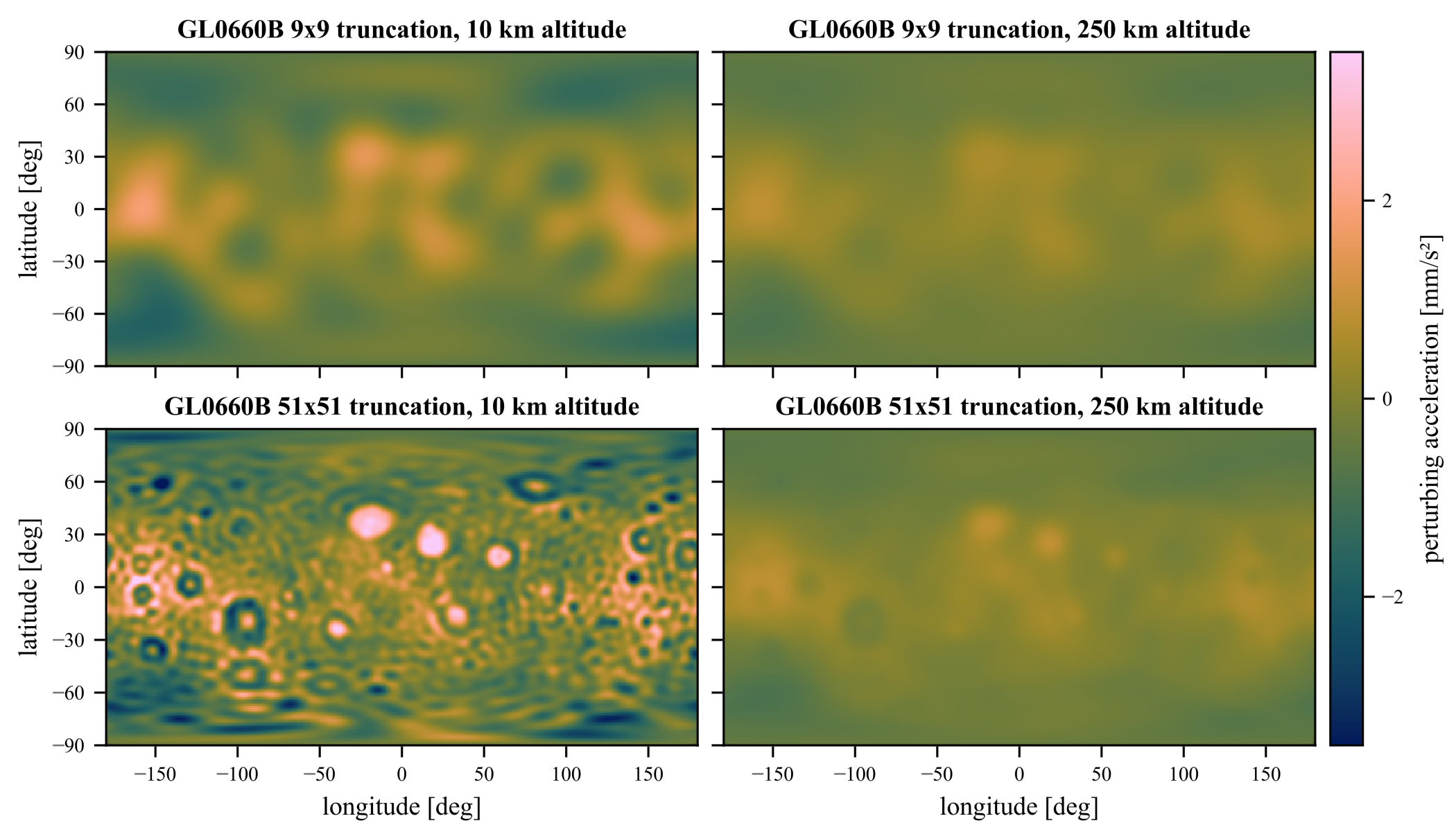}
\caption{Gravitational perturbations for 9x9 and 51x51 gravitational potentials at $10$ and $250$ km above the equatorial radius of the Moon, $1737.4 \si{km}$.}
\label{fig_harmonics}
\end{figure}

This choice is difficult and is mission dependent because of the varying impacts that altitude has on the effects of the spherical harmonic perturbations to Keplerian gravity. This is demonstrated in Fig.~\ref{fig_harmonics}, wherein the perturbations arising from the non-linear lunar gravity field are plotted for different truncation levels and altitudes. Whilst it is apparent that the $9\times9$ model does capture some of the perturbation behavior, particularly at the $250$ km altitude, at lower altitudes, the differences become particularly apparent. 

A straightforward analysis was conducted on the nominal BINAR mission, which should be applicable to all three test cases as it exhibits the lowest orbit altitude ranges. This analysis considered different gravity models of order and degree ranging from $3\times3$ to $81\times81$, and then compared the propagation of the eccentricity vector (as it is important to the translation theorem) to a mission with order and degree $101\times101$. In order to remove longitude-dependent effects resulting from the starting configuration, the longitude of ascending node is varied in 10 steps from $\deg{0}$ to $\deg{360}$ and the results averaged.

\begin{figure}[h]
\centering
\includegraphics[width=3.25in]{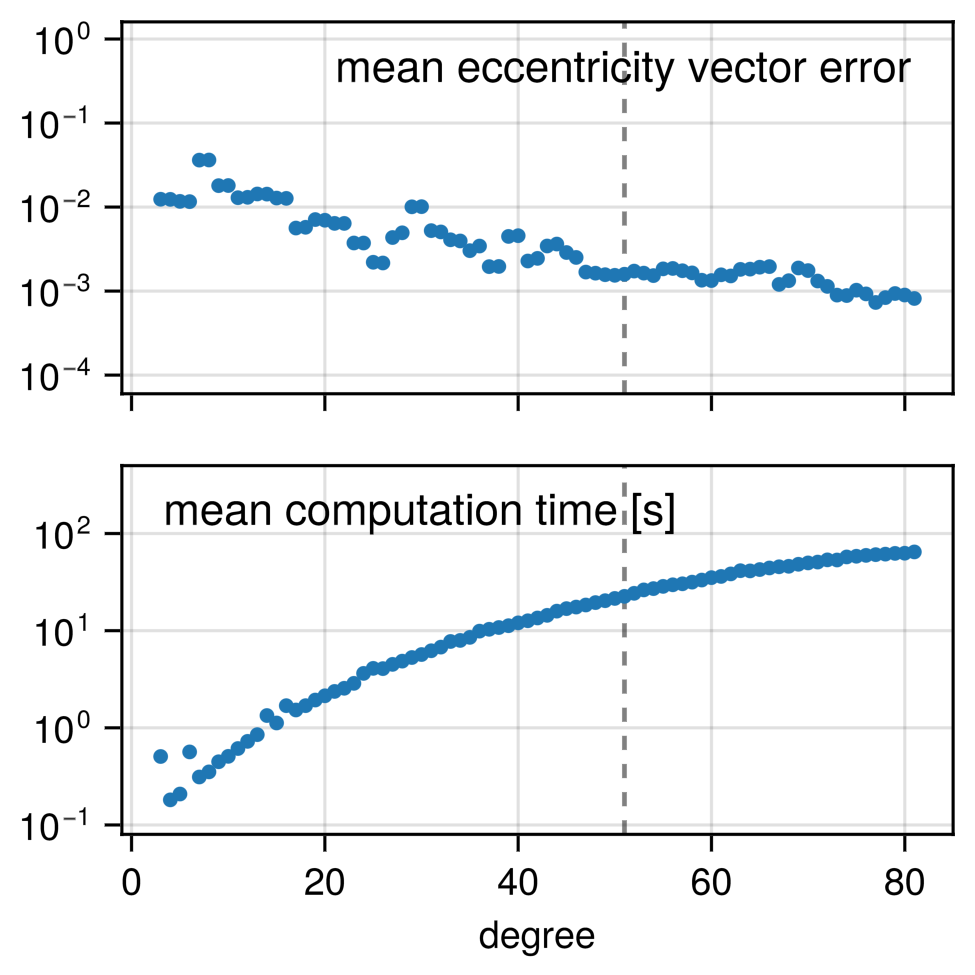}
\caption{Eccentricity vector errors and computation times for different truncation orders for the nominal BINAR case.}
\label{figure_truncation_justification}
\end{figure}

The results of this analysis are observed in Fig.~\ref{figure_truncation_justification}. It is apparent that as the model order and degree increase, the mean eccentricity vector error decreases while the mean computation time increases in a manner similar to the expected $\mathcal{O}(n^2)$. Additionally, it is clear that odd-degree harmonics are primarily responsible for changes in eccentricity vector error. 

The dotted line represents the selected truncation order and degree of $51\times51$, which was chosen because it appears to be where improvements in the mean eccentricity vector error appear to stabilize and still allow reasonable computational times of approximately 15 seconds (for 90 days of propagation) on a single thread of an AMD Ryzen 7 5800X3D processor. In general, it is worth noting the agreement of this truncation with previous analyses from the literature based on the calculation of frozen orbits \citep{Konoplivetal1994, LaraFerrerSaedeleer2009}.

\section{\label{section_translation}Justification of the Translation Theorem}
The time dependence of the lunar gravitational potential is avoided within a rotating frame attached to the surface of the Moon. In such a frame, the dynamics become conservative, which allows for the problem to be advantageously approached by canonical perturbation theory. Specifically, the dynamics stemming from the non-spherical lunar gravitational potential are obtained from a Hamiltonian
\citep{LaraLopezPerezSanJuan2020}: 
\begin{equation} \label{equation_hamiltonian}
\mathcal{H}=-\frac{\mu}{2a}\Big[1+2\frac{\dot{\theta}}{n}\sqrt{1-e^2}\cos{I}
+2\frac{a^2\eta}{r^2}\sum_{i\ge2}\sum_{j=0}^{i}V_{i,j}\Big],
\end{equation}
where $\mu$ is the gravitational parameter of the Moon, $\dot{\theta}$ is the mean rotation rate of the central body, ($a$, $e$, $I$, $\Omega$, $\omega$, $f$) are the usual Keplerian orbital elements, $n=\sqrt{\mu/a^3}$ is the mean motion of the satellite, $\eta=\sqrt{1-e^2}$, $r=a\eta^2/(1+e\cos{f})$ is the distance from the center of mass of the Moon, and $\nu=\Omega-\dot{\theta}{t}$ is the longitude of the ascending node in the fixed frame of the Moon, with $t$ denoting the time. The functions $V_{i,j}$ are defined as \citep{Lara2021}, 
\begin{align} \nonumber
V_{i,j} =&\; \frac{\alpha^i}{p^i}\eta\sum_{k=0}^{i}{F}_{i,j,k}
\sum_{m=0}^{i-1}\binom{i-1}{m}\frac{e^m}{2^m}\sum_{l=0}^m\binom{m}{l}
\left[\left(S_{i,j}\cos\psi_{i-2k,j}-C_{i,j}\sin\psi_{i-2k,j}\right) \right. \\ \label{equation_v_ij}
& \left. \times\sin(i-2k-m+2l)f
+\left(C_{i,j}\cos\psi_{i-2k,j}+S_{i,j}\sin\psi_{i-2k,j}\right)\cos(i-2k-m+2l)f \right],
\end{align}
in which the arguments $\psi_{i-2k,j}$ are computed from the definition
\begin{equation} \label{equation_psi}
\psi_{l,m}=l\omega+m\nu-(l-m)_\pi,
\end{equation}
where $(l-m)_\pi\equiv\frac{\pi}{2}[(l-m)\bmod2]$ is a parity correction, $S_{i,j}$ and $C_{i,j}$ denote the harmonic coefficients of the lunar gravitational potential, $p=a\eta^2$ is the semilatus rectum, and $F_{i,j,k}$ are Kaula's inclination functions \citep{Kaula1966} which, after \citep{Izsak1964} are more efficiently evaluated in terms of the half-inclination angle \citep{Allan1965,Allan1973,Gooding1992}.

Note that whilst Equation~\eqref{equation_hamiltonian} is displayed in terms of the usual Keplerian elliptic elements for engineering insight, these elements must be considered as functions of some set of canonical variables for the whole Hamiltonian perturbation treatment. For example, the canonical Delaunay variables $\ell=M$, $g=\omega$, $h=\nu$, $L=\sqrt{\mu{a}}$, $G=L\eta$, and $H=G\cos{I}$. 

Moreover, \gls{eLLO} clearly stay away from deep tesseral resonances. Indeed, on account of the lunar gravitational parameter of $\mu=4902.8\,\mathrm{km^3/s^2}$, \glspl{eLLO} have periods shorter than, but close to 2 hours. This is just a few thousandths of the lunar sidereal period of about 27.3 days and combined with the fact that the lunar harmonic coefficients are $\mathcal{O}(10^{-4})$ or smaller, this clearly confers a perturbation arrangement to the Hamiltonian~\eqref{equation_hamiltonian}. 

\begin{figure}[t]
\centering
\includegraphics[width=\textwidth]{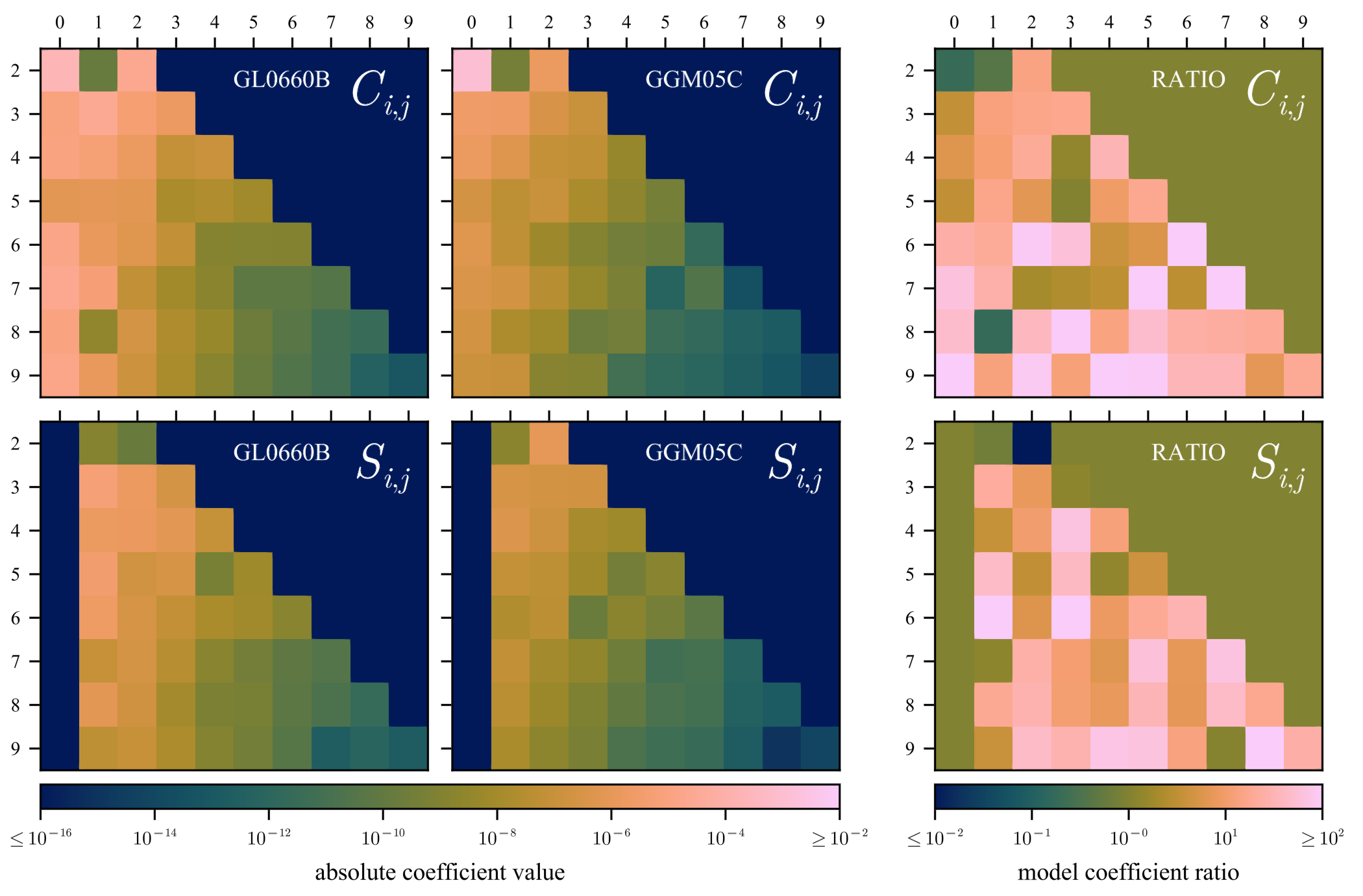}
\caption{\label{fig:gravity_model_coefficients} Comparing the gravity model coefficients between the lunar GL0660B model \citep{konoplivJPLLunarGravity2013} and the terrestial GGM05C model \citep{riesDevelopmentEvaluationGlobal2018}.}
\label{figure_harmonic_coefficients}
\end{figure}

Fig.~\ref{figure_harmonic_coefficients} demonstrates the differences between the lunar harmonic coefficients and those of the Earth. It is clear that the overwhelming dominance of the second zonal harmonic ($C_{2,0}$) does not occur in the case of the Moon; in fact, it is more than 5 times smaller. In contrast, most of the other harmonic coefficients are more than 10 times larger for the Moon compared to Earth. Consequently, the lumpy gravitational field of the Moon makes the consideration of higher-degree truncations of the lunar potential unavoidable for realistic purposes, unlike that of the Earth \citep{Konoplivetal1994,LaraFerrerSaedeleer2009,LaraSaedeleerFerrer2009}.

\subsection{\label{section_averaging_mean}Averaging the mean anomaly}
Canonical perturbation theory is used for the removal of short-period effects from the Hamiltonian \eqref{equation_hamiltonian}, such as those originating from the mean anomaly. Due to the relatively small size of all the harmonic coefficients of the lunar gravitational potential, a linear theory such as the popular choice of \citep{Kaula1966} is sufficient for our purposes.

The highest frequencies of the motion, which are associated with the fast advance of the mean anomaly, are filtered first. For this purpose, the Hamiltonian \eqref{equation_hamiltonian} is written in the form
\begin{equation} \label{equation_hamiltonian_perturbation}
\mathcal{H}=\mathcal{H}_0+\epsilon\mathcal{H}_1,
\end{equation}
where 
\begin{align} \label{equation_hamiltonian_0}
\mathcal{H}_0=&-\frac{\mu}{2a}, \\ \label{equation_hamiltonian_1}
\mathcal{H}_1=&-H\dot{\theta}-\frac{\mu}{a}\frac{a^2\eta}{r^2}\sum_{i\ge{2}}\sum_{j=0}^{i}V_{i,j},
\end{align}
in which $H=\sqrt{\mu{a}}\eta\cos{I}$ is the third component of the specific angular momentum vector, and the small parameter $\epsilon$ indicates that the perturbation solution will apply to regions in which $|\mathcal{H}_1|\ll|\mathcal{H}_0|$. Because $-H\dot{\theta}=(\mu/a)(\dot{\theta}/n)\eta\cos{I}$, this essentially means that $\dot{\theta}/n$ is much smaller than unity---just as is the case of \glspl{eLLO}.

Up to the order of $\epsilon$, the canonical transformation that removes the mean anomaly $M=\ell$ is obtained from the generating function \citep{Deprit1969, Lara2021},
\begin{equation} \label{generating_function_w}
\mathcal{W}=\frac{1}{n}\int(\mathcal{H}_1-\mathcal{H}^*_1)\,\mathrm{d}M.
\end{equation}
To avoid the appearance of secular terms in the transformation, the term $\mathcal{H}^*_1$ is chosen by averaging Equation~\eqref{equation_hamiltonian_1} over the mean anomaly. From the standard relations of the ellipse we obtain $\eta\,\mathrm{d}M=(r/a)^2\mathrm{d}f$, and hence we choose $\mathcal{H}^*_1=\langle\mathcal{H}_1\rangle_M=\langle\mathcal{H}_1r^2/(a^2\eta)\rangle_f$. That is,
\begin{equation} \label{equation_hamiltonian_1_averaged}
\langle\mathcal{H}_1\rangle_M=-H\dot{\theta}
-\frac{\mu}{a}\sum_{i\ge{2}}\sum_{j=0}^{i}\langle{V}_{i,j}\rangle_f,
\end{equation}
in which
\begin{equation} \label{equation_v_ij_averaged}
\langle{V}_{i,j}\rangle_f=\frac{\alpha^i}{a^i}\frac{1}{\eta^{2i-1}}\sum_{k=1}^{i-1}{F}_{i,j,k}{G}_{i,k}\left(C_{i,j}\cos\psi_{i-2k,j}+S_{i,j}\sin\psi_{i-2k,j}\right)
\end{equation}
is readily obtained by constraining the innermost summation of Equation~\eqref{equation_v_ij} to the value $l=\frac{1}{2}(m-i)+k$. The functions $G_{i,k}$ in Equation~\eqref{equation_v_ij_averaged} are obtained by multiplying the corresponding closed-form eccentricity functions of Kaula ($G_{i,k,2k-1}$) by $\eta^{2i-1}$. Recall that ${G}_{i,k}$ is $\mathcal{O}(1)$ when $i=2k$, and at least $\mathcal{O}(e)$ otherwise.

By substituting Equations~\eqref{equation_hamiltonian_1} and \eqref{equation_hamiltonian_1_averaged} into Equation~\eqref{generating_function_w}, we obtain
\begin{equation} \label{generating_function_w_expanded}
\mathcal{W}=-G\sum_{i\ge{2}}\sum_{j=0}^{i}\Big(\frac{f-M}{\eta}\langle{V}_{i,j}\rangle_f+W_{i,j}\Big),
\end{equation}
in which $G=H/\cos{I}$ denotes the modulus of the specific angular momentum vector, and 
\begin{align} \nonumber
{W}_{i,j}=&\;
\frac{\alpha^i}{p^i}\sum_{k=0}^{i}{F}_{i,j,k}
\sum_{m=0}^{i-1}\binom{i-1}{m}\frac{e^m}{2^m}\sum_{\substack{l=0 \\ l\ne{l^*} } }^m\binom{m}{l} \\ \label{Wij}
&\times\frac{C_{i,j}\sin[2(l-l^*)f+\psi_{i-2k,j}] -S_{i,j}\cos[2(l-l^*)f+\psi_{i-2k,j}]}{2(l-l^*)},
\end{align}
where $l^*=k+\frac{1}{2}(m-i)$.

The mean anomaly averaging procedure ends after replacing the original variables with the new prime variables in the new Hamiltonian $\mathcal{H}^*=\mathcal{H}_0+\epsilon\mathcal{H}^*_1$, which shows that $L'$, and hence $a=L'^2/\mu$, are integrals of the transformed flow. The particular equations for the transformation can then be derived from the generating function if desired for the canonical variables or any function of them.

\subsection{\label{section_removing_node}Removing the argument of the node}
The previous filtering of the highest frequencies of the motion, i.e. the terms relating to the mean anomaly, still leaves the Hamiltonian with terms that depend on the longitude of the ascending node in the rotating frame $\nu=h'$, which modulate the long-term dynamics. The periods of these terms are short compared with the much slower evolution of the argument of the perilune, which commonly takes several years to complete a cycle \citep{BeckmanLamb2007}. Therefore, a second averaging is performed to additionally remove these short-period terms from the Hamiltonian by a new canonical transformation to double-prime variables.

We begin from the perturbation Hamiltonian $\mathcal{K}=\mathcal{K}_0+\varepsilon\mathcal{K}_1$, with the mean anomaly removed, where
\begin{align} \label{hamiltonian_K0_mean}
\mathcal{K}_0=&-\frac{\mu}{2a}-H'\dot{\theta},
\\ \label{hamiltonian_K1_mean}
\mathcal{K}_1=&-\frac{\mu}{a}\sum_{i\ge{2}}\sum_{j=0}^{i}\langle{V}_{i,j}\rangle_f.
\end{align}
Since $H=\mu\eta\cos{I}/({an})$, with this arrangement the new formal small parameter $\varepsilon$ indicates that the validity of the solution will be constrained to the region in which the lunar harmonic coefficients $C_{i,j}$ and $S_{i,j}$ (in Equation~\eqref{equation_v_ij}) are at least $\mathcal{O}(\dot{\theta}/n)^2$. Recall that now $a=L'^2/\mu$, and $\langle{V}_{i,j}\rangle_f$ is a function of the prime variables. Then, we split the perturbation term of the Hamiltonian as
\begin{equation} \label{hamiltonian_K1_mean_2}
\mathcal{K}_1=-\frac{\mu}{a}\sum_{i\ge{2}}\langle{V}_{i,0}\rangle_f
-\frac{\mu}{a}\sum_{i\ge{2}}\sum_{j=1}^{i}\sum_{k=1}^{i-1}T'_{i,j,k},
\end{equation}
thus explicitly separating the contribution of the zonal terms of Equation~\eqref{equation_v_ij_averaged}
\begin{equation} \label{equation_v_ij_averaged_zonal}
\langle{V}_{i,0}\rangle_f=C_{i,0}\frac{\alpha^i}{a^i}\frac{1}{\eta^{2i-1}}\sum_{k=1}^{i-1}{F}_{i,0,k}{G}_{i,k}\cos\psi_{i-2k,0},
\end{equation}
from the remaining tesseral and sectoral terms
\begin{equation} \label{Tpijk}
T'_{i,j,k}=\frac{\alpha^i}{a^i}{F}_{i,j,k}\frac{{G}_{i,k}}{\eta^{2i-1}}\left(C_{i,j}\cos\psi_{i-2k,j}+S_{i,j}\sin\psi_{i-2k,j}\right),
\end{equation}
where all of them depend on $\nu=h'$ because $j\ge1$ in Equation~\eqref{hamiltonian_K1_mean_2}.

The new transformation that removes the periodic terms related to the lunar rotation period is then derived from the new generating function 
\begin{equation}
\mathcal{W}'=\frac{1}{\dot{\theta}}\int(\mathcal{K}^*_1 - \mathcal{K}_1)\,\mathrm{d}\nu, \label{equation_generating_function_2}
\end{equation}
where, to guarantee that $\mathcal{W}'$ is purely periodic in $\nu=h'$, we select $\mathcal{K}^*_1=\langle\mathcal{K}_1\rangle_{h'}$. This choice removes all of the sectoral and tesseral terms from Equation~\eqref{hamiltonian_K1_mean_2}, yielding
\begin{equation} \label{hamiltonian_K1_mean_3}
\mathcal{K}^*_1=-\frac{\mu}{a}\sum_{i\ge{2}}\langle{V}_{i,0}\rangle_f,
\end{equation}
which only involves the effects of the zonal harmonics. Then, $\mathcal{W}'$ becomes defined as
\begin{equation} \label{generating_function_w_dash}
\mathcal{W}'=-L'\frac{n}{\dot{\theta}}\sum_{i\ge{2}}\sum_{j=1}^{i}\sum_{k=1}^{i-1}T_{i,j,k},
\end{equation}
where
\begin{equation} \label{function_tijk}
T_{i,j,k}=\int{T}'_{i,j,k}\,\mathrm{d}h'=\frac{\alpha^i}{a^i}{F}_{i,j,k}\frac{{G}_{i,k}}{\eta^{2i-1}}\frac{C_{i,j}\sin\psi_{i-2k,j}-S_{i,j}\cos\psi_{i-2k,j}}{j}.
\end{equation}

The procedure ends after writing $\mathcal{K}^*=\mathcal{K}_0+\varepsilon\mathcal{K}^*_1$ in the new, double-prime variables, which shows that $H''$, and hence the mean inclination of the circular orbits $I_\mathrm{circular}=\arccos(H''/L'')$ is a formal integral of the Hamiltonian. It is worth noting that the term $n/\dot{\theta}=\mathcal{O}(10^2)$, which scales the generating function of the periodic corrections in Equation~\eqref{generating_function_w_dash}, produces large monthly oscillations of the eccentricity that cannot be ignored in the design of low-lunar orbits \citep{BeckmanLamb2007,holtExtremelyLowaltitudeLunar2024}.

\subsection{\label{section_long_term}Long-term dynamics of the eccentricity vector}
After the double averaging process, the new Hamiltonian is free from short-period terms and accurate to $\mathcal{O}(\dot{\theta}/n)^2$. It takes the form
\begin{equation} \label{double_averaged_hamiltonian}
\mathcal{L}=-\frac{\mu}{2a} -H''\dot{\theta} -\frac{\mu}{a}\sum_{i\ge{2}}\langle{V}_{i,0}\rangle_f,
\end{equation}
where $a$ and $\langle{V}_{i,0}\rangle_f$ are now functions of the double-prime variables. Equation~\eqref{double_averaged_hamiltonian} is a one-degree-of-freedom Hamiltonian $\mathcal{L}=\mathcal{L}(g'',G'')$ in the mean argument of periapsis $\omega=g''$ and the mean total angular momentum $G''$, whose long-term variations are obtained from the Hamilton equations. Alternatively, due to the indeterminacy of the argument of periapsis $\omega$ in the case of circular orbits, the reduced flow is preferably described by the semi-equinoctial elements that define the components of the eccentricity vector in the nodal frame. Namely,
\begin{equation}
\mathcal{C}=e\cos\omega,\qquad \mathcal{S}=e\sin\omega,
\end{equation}
where $e=e(G'')\equiv\sqrt{1-(G''/L'')^2}$. The corresponding variation equations are then ${\mathrm{d}\mathcal{C}}/{\mathrm{d}t}=\{\mathcal{C},\mathcal{L}\}$ and ${\mathrm{d}\mathcal{S}}/{\mathrm{d}t}=\{\mathcal{S},\mathcal{L}\}$ where curly brackets denote the Poisson brackets operator. That is,
\begin{equation} \label{eccentricity_vector_equations}
\frac{\mathrm{d}\mathcal{C}}{\mathrm{d}t}=\frac{\partial{\mathcal{C}}}{\partial{g''}}\frac{\partial\mathcal{L}}{\partial{G''}}
-\frac{\partial{\mathcal{C}}}{\partial{G''}}\frac{\partial\mathcal{L}}{\partial{g''}}, \qquad 
\frac{\mathrm{d}\mathcal{S}}{\mathrm{d}t}=\frac{\partial{\mathcal{S}}}{\partial{g''}}\frac{\partial\mathcal{L}}{\partial{G''}}
-\frac{\partial{\mathcal{S}}}{\partial{G''}}\frac{\partial\mathcal{L}}{\partial{g''}},
\end{equation}
which, after computation, must be formulated in terms of $\mathcal{C}$ and $\mathcal{S}$. This requires the replacement of the appearance of $e$ and $\omega$ in the long-term potential $\langle{V}_{i,0}\rangle_f$ in terms of $\mathcal{C}$ and $\mathcal{S}$. It is worth noting that alternative derivations of the mean eccentricity vector dynamics can be found elsewhere \citep{Cook1992,Lara2023BRICS}.

Firstly, we note from Equation~\eqref{equation_psi} that $\psi_{i-2k, 0}=(i-2k)\omega - \frac{\pi}{2}[i\bmod2]$. Hence,
\begin{equation}
\cos\psi_{i-2k,0} = \beta_{i-2k}^{i*} \sum_{l=0}^{\frac{|i-2k|-i^*}{2}} \left[ (-1)^l \binom{|i - 2k|}{2l + i^*} \sin^{2l+i^*} \omega \cos^{|i-2k|-2l-i^*} \omega \right ],
\end{equation}
where $\beta_{i-2k}=\text{sign}(i-2k)$ and $i^*=i\bmod2$. Then, Equation~\eqref{equation_v_ij_averaged} becomes
\begin{equation}
    \langle V_{i,0} \rangle_f = C_{i,0} \frac{\alpha_i}{a^i} \frac{1}{\eta^{2i-1}} \sum_{k=1}^{i-1} F_{i,0,k}(I) \widetilde{G}_{i,k}(\mathcal{C}, \mathcal{S}) Q_{i,k}(\mathcal{C}, \mathcal{S}),
\end{equation}
in which we abbreviated $\widetilde{G}_{i,k}=G_{i,k}/(\mathcal{C}^2+\mathcal{S}^2)^{\left|i-2k\right|/2}$
\begin{equation}
    Q_{i,k} = \beta_{i-2k} \sum_{t=0}^{\frac{|i-2k|-i^*}{2}} \left[ (-1)^t \binom{|i-2k|}{2t + i^*} \mathcal{C}^{|i-2k| -2t-i^*} \mathcal{S}^{2t+i^*} \right].
\end{equation}
Therefore, $\langle V_{i,0} \rangle_f = \langle V_{i,0} \rangle_f (I, \mathcal{C}, \mathcal{S}; a)$ and the double averaged Hamiltonian~\eqref{double_averaged_hamiltonian} is written explicitly as $\mathcal{L}=\mathcal{L}(I, \mathcal{C}, \mathcal{S}; a, H'')$. 

Then, taking the requisite partial derivatives, a sequence of operations can be taken to show that the mean variations of the eccentricity vector components turn into
\begin{align}
    \frac{\partial \mathcal{C}}{\partial t} &= \frac{1}{L}\left[\eta \frac{\partial \mathcal{L}}{\partial \mathcal{S}} + \mathcal{S} \left(\frac{\eta}{e} \frac{\partial \mathcal{L}}{\partial e} - \frac{\partial \mathcal{L}}{\partial \eta} - \frac{c}{s\eta} \frac{\partial \mathcal{L}}{\partial I}\right) \right], \\
    \frac{\partial \mathcal{S}}{\partial t} &= -\frac{1}{L}\left[\eta \frac{\partial \mathcal{L}}{\partial \mathcal{C}} + \mathcal{C} \left(\frac{\eta}{e} \frac{\partial \mathcal{L}}{\partial e} - \frac{\partial \mathcal{L}}{\partial \eta} - \frac{c}{s\eta} \frac{\partial \mathcal{L}}{\partial I}\right) \right],
\end{align}
which, after substitution of the requisite partial derivatives and conversion to the time scale $\text{d}\tau = n \,\text{d}t$, leaves
\begin{align}
    \frac{\partial \mathcal{C}}{\partial \tau} &= -\sum_{i\geq 2} C_{i,0} \frac{\alpha^i}{p^i} \sum_{k=1}^{i-1}\left[ \frac{\partial Q_{i,k}}{\partial \mathcal{S}} \left(\eta^2 F_{i,0,k} \widetilde{G}_{i,k} \right) + \mathcal{S} \left(\eta^2 F_{i,0,k} G_{i,k}^\dagger + F_{i,j,k}^\dagger \widetilde{G}_{i,k} \right) Q_{i,k}\right], \label{equation_mean_motion_1} \\
    \frac{\partial \mathcal{S}}{\partial \tau} &= \sum_{i\geq 2} C_{i,0} \frac{\alpha^i}{p^i} \sum_{k=1}^{i-1}\left[ \frac{\partial Q_{i,k}}{\partial \mathcal{C}} \left(\eta^2 F_{i,0,k} \widetilde{G}_{i,k} \right) + \mathcal{C} \left(\eta^2 F_{i,0,k} G_{i,k}^\dagger + F_{i,j,k}^\dagger \widetilde{G}_{i,k} \right) Q_{i,k}\right], \label{equation_mean_motion_2}
\end{align}
in which we have further abbreviated
\begin{align}
    G_{i,k}^\dagger &= \frac{G_{i,k}'}{e^{1+|i-2k|}} - \frac{|i-2k|}{e^2}\widetilde{G}_{i,k}, \\
    F_{i,j,k}^\dagger &= (2i-1)F_{i,j,k} - F_{i,j,k}'\cot I.
\end{align}

The arrangement of Equations~\eqref{equation_mean_motion_1} and \eqref{equation_mean_motion_2} shows that the evolution of the eccentricity vector is free of singularities for circular orbits, since $I=\arccos H''/G'', G''=L''\eta, \eta=\sqrt{1-e^2}, e=\sqrt{
\mathcal{C}^2+\mathcal{S}^2}$, and $H''$ and $L''$ are formal integrals.

Generally, the mean dynamics of the eccentricity vector must be integrated numerically. However, the equilibria of Equations~\eqref{equation_mean_motion_1} and \eqref{equation_mean_motion_2}, which give rise to frozen orbits, can be obtained from algebraic equations. Among these equations, it can be shown that the variation of $\mathcal{S}$ vanishes for $\mathcal{C}=0$, which means that the fixed points appear when $\omega=\pm\frac{1}{2}\pi$ and hence $\mathcal{S} = \pm e$. In that case, the vanishing of the variation in $\mathcal{S}$ yields
\begin{equation}
    0=\sum_{i\geq 2} C_{i,0} \frac{\alpha^i}{p^i} \cos \frac{\pi}{2} (i \mp i^*)\sum_{k=1}^{i-1}(-1)^k \left[ 
    G_{i,k}F_{i,0,k}' \cot I - \left( \frac{\eta^2}{e} G_{i,k}'  + (2i - 1)G_{i,k}\right) F_{i,0,k}
    \right], \label{equation_mean_motion_fixed}
\end{equation}
where $i^*=i \bmod 2$. The minus sign in $\cos \frac{\pi}{2} (i\mp i^*)$ corresponds to the case where $\omega=\frac{1}{2}\pi$, and the plus sign to the opposite. Since this is an implicit equation, in general, it must be approached numerically using root-finding methods to find solutions.

\begin{figure}[!ht] 
\centering
\includegraphics[width=\textwidth]{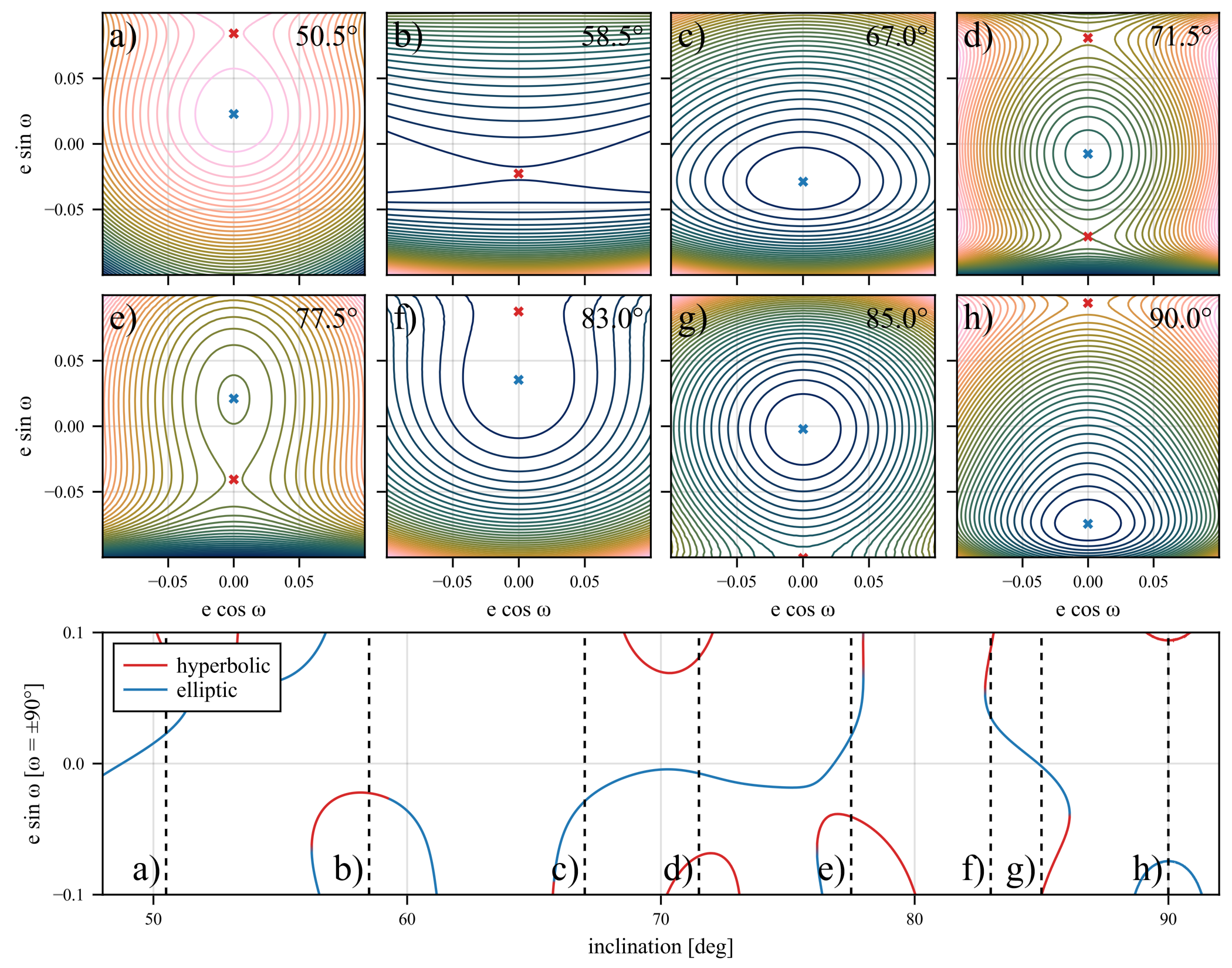}
\caption{Mean eccentricity vector dynamics for higher inclinations ($50$ km altitude, $51\times 0$ lunar potential).}
\label{figure_equipotentials_51}
\end{figure}

The utility of this approach in the construction and analysis of high-inclination orbits is demonstrated in Fig.~\ref{figure_equipotentials_51}. The contour plots visualize the structure of the non-linear flow about specific orbits of the doubly averaged Hamiltonian~\eqref{double_averaged_hamiltonian}. Subsequently, the bottom panel shows a bifurcation analysis across inclinations on the mean motion dynamics for low eccentricities created using Equation~\eqref{equation_mean_motion_fixed}. It is worth remarking that the correspondence between the inclination-eccentricity diagram and contour plots is not strict, as, more precisely, the inclination of the former refers to that of the frozen orbit which is slightly different from the value $I_\text{circular}$ labeling the eccentricity vector plots. This is due to the integral relation $\cos I_\text{circular} = H''/L'' = \sqrt{1-e^2}\cos I$ which only implies $I_\text{circular}=I$ when $e=0$.

The stability structure of the bifurcation diagram is complex, with both Hamiltonian-Hopf (elliptic/hyperbolic points spontaneously changing stability) and Saddle-Center (elliptic and hyperbolic points colliding/appearing spontaneously) bifurcations present. It is immediately apparent from this analysis that only a few inclinations are predicted to permit stable (elliptic) frozen orbits with near-circular eccentricity. These are at approximately $\ang{58}$, $\ang{71}$, $\ang{76}$, and $\ang{85}$, for the $51\times0$ lunar potential. Plots a), d), e), and g) show the mean motion flow near these inclinations, which each show a clear locally circular structure around the central elliptic point. The practical use of these orbits may be degraded when there are nearby hyperbolic points which modulate the local flow, the effects of which are most clearly seen in plots e) and f), where orbits which do not very closely follow the central elliptic point will be clearly modulated by the nearby hyperbolic point.

It should be noted that the effects of the non-spherical lunar potential become stronger for lower orbits and weaker for higher orbits, which means that the map of stable frozen orbits collapses into just a few narrow strips of inclination in the first case and expands to support a wider range of inclinations in the second case \citep{Konoplivetal1994}. 

Finally, once the mean dynamics of the eccentricity vector are solved, the variations of the mean values of the mean anomaly $\ell''$ and the argument of the node in the rotating frame $h''$ are integrated from the corresponding Hamilton equations $d\ell''/dt = \partial \mathcal{L}/ \partial{L''}$ and $dh''/dt = \partial \mathcal{L}/\partial H''$. For the latter, in the new time scale $\tau$, we readily obtain
\begin{equation}
    \frac{dh''}{dt}=-\frac{\dot{\theta}}{n} + \sum_{i\geq 2} C_{i, 0} \frac{\alpha^i}{p^i} \sum_{k=1}^{i-1}\frac{1}{s}F_{i, 0, k}' \widetilde{G}_{i, k} Q_{i, k}. \label{equation_mean_argument}
\end{equation}

\subsection{\label{section_m_monthly}Monthly perturbations and the translation theorem}
Although the long-term dynamics are correctly captured by the zonal part of the lunar potential, the mean evolution of the eccentricity vector alone may not be significant enough for the purpose of mission designing of low-lunar orbits. Indeed, the large amplitude of the periodic terms due to the sectoral and tesseral monthly contributions may highly perturb the orbits of interest much more significantly than would be expected from the long-term mean eccentricity vector evolution. Therefore, these terms cannot be ignored for purposes of mission design, and the first-order corrections resulting from the generating function $\mathcal{W}'$ in Equation~\eqref{equation_generating_function_2} must be taken into account.

In particular, the periodic, m-monthly corrections $\Delta{\mathcal{C}}$ and $\Delta{\mathcal{S}}$, to the eccentricity vector are
\begin{align}
\Delta{\mathcal{C}}=\{e\cos\omega,\mathcal{W}'\}=&\; -e\sin\omega\frac{\partial\mathcal{W}'}{\partial{G}'}
+\cos\omega\frac{\eta^2}{eL}\frac{\partial\mathcal{W}'}{\partial{g}'}, \label{equation_e1_formula}\\
\Delta{\mathcal{S}}=\{e\sin\omega,\mathcal{W}'\}=&\; e\cos\omega\frac{\partial\mathcal{W}'}{\partial{G}'}
+\sin\omega\frac{\eta^2}{eG}\frac{\partial\mathcal{W}'}{\partial{g}'},
\end{align}
where, according to Equations~\eqref{generating_function_w_dash} and \eqref{function_tijk},
\begin{align}
\frac{\partial\mathcal{W}'}{\partial{g}'}=& \;
L'\frac{n}{\dot{\theta}}\sum_{i\ge{2}}\frac{\alpha^i}{a^i}\sum_{j=1}^{i}\sum_{k=1}^{i-1}
{F}_{i,j,k}\frac{{G}_{i,k}}{\eta^{2i-1}}
\frac{C_{i,j}\cos\psi_{i-2k,j}+S_{i,j}\sin\psi_{i-2k,j}}{j}(i-2k), \\
\frac{\partial\mathcal{W}'}{\partial{G}'}=& \;
L'\frac{n}{\dot{\theta}}\sum_{i\ge{2}}\frac{\alpha^i}{a^i}\sum_{j=1}^{i}\sum_{k=1}^{i-1}\frac{C_{i,j}\sin\psi_{i-2k,j}-S_{i,j}\cos\psi_{i-2k,j}}{j}\frac{\partial}{\partial{G}'}\Big({F}_{i,j,k}\frac{{G}_{i,k}}{\eta^{2i-1}}\Big),
\end{align}
and
\begin{align*}
\frac{\partial}{\partial{G}'}\left({F}_{i,j,k}\frac{{G}_{i,k}}{\eta^{2i-1}}\right)=
\frac{1}{G'\eta^{2i-1}}\left[ 
\left(F'_{i,j,k}\cot{I}-(2i-1)F_{i,j,k}\right)G_{i,k}
-{F}_{i,j,k}\frac{\eta^2}{e}G'_{i,k} \right].
\end{align*}

For the m-monthly corrections to $\mathcal{S}$, $\Delta \mathcal{S}$, that is,
\begin{align} \nonumber
\Delta{\mathcal{S}}=&\; \frac{n}{\dot{\theta}}\sum_{i\ge{2}}\frac{\alpha^i}{a^i}\frac{1}{\eta^{2i-2}}\sum_{j=1}^{i}\sum_{k=1}^{i-1}\left\{
(i-2k)\frac{{G}_{i,k}}{e}{F}_{i,j,k}
\frac{C_{i,j}\cos\psi_{i-2k,j}+S_{i,j}\sin\psi_{i-2k,j}}{j}\sin\omega \right. \\ \label{moonS01}
&\left. +\left[ \left(\cot{I}F'_{i,j,k}
-(2i-1){F}_{i,j,k}\right)\frac{e{G}_{i,k}}{\eta^2}
-{F}_{i,j,k}G'_{i,k} \right]
\frac{C_{i,j}\sin\psi_{i-2k,j}-S_{i,j}\cos\psi_{i-2k,j}}{j}\cos\omega \right\}.
\end{align}
Divisions by the eccentricity stemming from the term $G_{i,k}/e$ in the first row of Equation~\eqref{moonS01} do not exist in fact due to the factor $i-2k$ multiplying this kind of term.

Next, taking into account the trigonometric identities
\begin{align*}
 2\sin\omega\sin\psi_{i-2k,j}=&\; (-1)^{(i-j)^\star}\left(\sin\psi_{i-2k+1,j}-\sin\psi_{i-2k-1,j}\right), \\
 2\sin\omega\cos\psi_{i-2k,j}=&\; (-1)^{(i-j)^\star}\left(\cos\psi_{i-2k+1,j}-\cos\psi_{i-2k-1,j}\right), \\
 2\cos\omega\sin\psi_{i-2k,j}=&\; (-1)^{(i-j)^\star}\left(\cos\psi_{i-2k+1,j}+\cos\psi_{i-2k-1,j}\right), \\
-2\cos\omega\cos\psi_{i-2k,j}=&\; (-1)^{(i-j)^\star}\left(\sin\psi_{i-2k+1,j}+\sin\psi_{i-2k-1,j}\right),
\end{align*}
where $(i-j)^\star=(i-j)\bmod2$, straightforward computations allow us to rearrange Equation~\eqref{moonS01} as
\begin{align*}
\Delta{\mathcal{S}}=&\; \frac{n}{\dot{\theta}}\sum_{i\ge{2}}\frac{\alpha^i}{a^i}\frac{i-1}{4}\sum_{j=1}^{i}\frac{(-1)^{(i-j)^\star}}{j}\sum_{k=1}^{i-1}
\binom{1}{\frac{i+1}{2}-k}{F}_{i,j,k}\left[
(i-2k-1)\left(S_{i,j}\sin\psi_{i-2k+1,j} 
\right.\right. \\& \left.\left. 
+C_{i,j}\cos\psi_{i-2k+1,j}\right)-(i-2k+1)\left(S_{i,j}\sin\psi_{i-2k-1,j}+C_{i,j}\cos\psi_{i-2k-1,j}\right)
+\mathcal{O}(e) \right].
\end{align*}
To this order of approximation, the binomial coefficient vanishes except when $k=\frac{1}{2}(i\pm1)$, in which case it takes the value 1, thus only involving the contribution of the non-zonal harmonics of odd degree in the main term of the periodic corrections to $\mathcal{S}$. That is, 
\begin{equation} \label{moonS01e0}
\Delta{\mathcal{S}}=-\frac{n}{\dot{\theta}}\sum_{i\ge 3}\frac{\alpha^i}{a^i}\frac{i-1}{2}\sum_{j=1}^{i}
\left[ \mathcal{S}_{i,j} +\mathcal{O}(e) \right],
\end{equation}
in which, for $i\ge3$ odd, and replacing $\psi_{0,j}=j\nu-j_\pi$ from Equation~\eqref{equation_psi},
\begin{equation} \label{moonSije0}
\mathcal{S}_{i,j}=(-1)^{(i-j)^\star}\left({F}_{i,j,\frac{i-1}{2}}+{F}_{i,j,\frac{i+1}{2}}\right)\frac{S_{i,j}\sin(j\nu-j_\pi)+C_{i,j}\cos(j\nu-j_\pi)}{j},
\end{equation}
with the parity correction $j_\pi=\frac{\pi}{2}(j\bmod2)$.

Analogous computations for Equation~\eqref{equation_e1_formula} result in
\begin{equation} \label{moonC01e0}
\Delta{\mathcal{C}}=\frac{n}{\dot{\theta}}\sum_{i\ge3}\frac{\alpha^i}{a^i}\frac{i-1}{2}\sum_{j=1}^{i}
\left[\mathcal{C}_{i,j}+\mathcal{O}(e)\right]
\end{equation}
in which, for $i\ge3$ odd,
\begin{equation} \label{moonCije0}
\mathcal{C}_{i,j}=(-1)^{(i-j)^\star}\left({F}_{i,j,\frac{i-1}{2}}-{F}_{i,j,\frac{i+1}{2}}\right)\frac{S_{i,j}\cos(j\nu-j_\pi)-C_{i,j}\sin(j\nu-j_\pi)}{j}.
\end{equation}

Remember that the inclination is coupled with the eccentricity through the formal integral $H''=G''\cos{I}=L''\eta\cos{I}$. However, because $\eta=1+\mathcal{O}(e^2)$, in the order of approximation of Equations~\eqref{moonS01e0} and \eqref{moonC01e0} we can safely use the inclination of the circular orbits $I_\mathrm{circular}=\arccos(H''/L'')$ in Equations~\eqref{moonSije0} and \eqref{moonCije0}.

Therefore, the bulk of the tesseral corrections of the lower lunar orbits neither depend on the argument of the perilune nor on the eccentricity, and are driven by $j$-monthly corrections. Consequently, starting from a given mean $a$, $I$ and $\nu$, the eccentricity vector will follow the same perturbation pattern irrespective of the starting point. This feature was soon recognized by mission designers of low-lunar orbits, who dubbed it the ``translation theorem'' and skillfully took advantage of it in their optimization procedures \cite{BeckmanLamb2007,holtExtremelyLowaltitudeLunar2024}.

\subsection{\label{section_monthly_pattern}Composition of mean motion and monthly perturbations}

\begin{figure}[!ht]
\includegraphics[width=\textwidth]{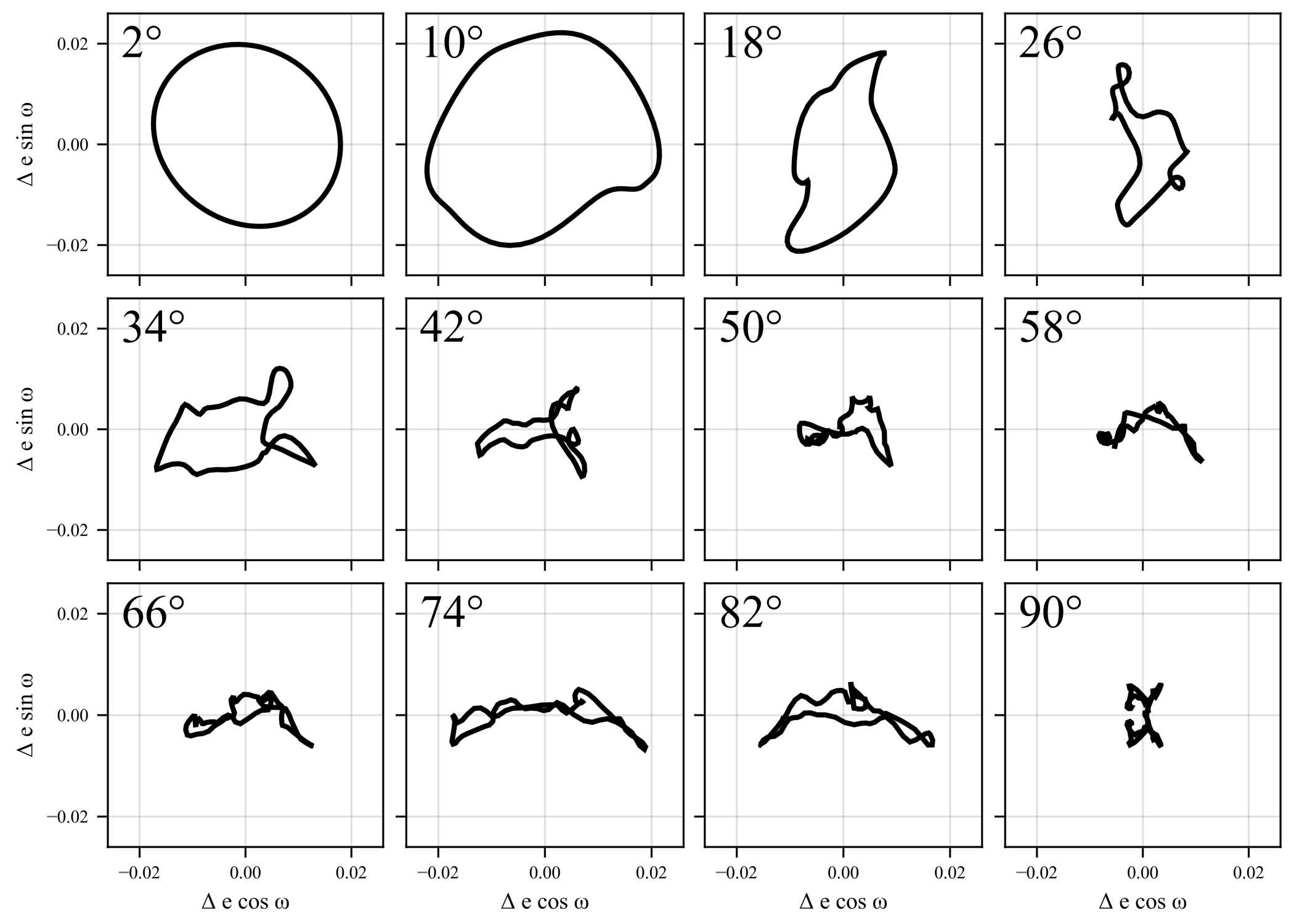}
\caption{Degree 51 correction patterns with different inclinations for an orbit similar to the LRO case.}
\label{figure_correction_patterns_inclination}
\end{figure}

To illustrate the significant influence of monthly perturbation patterns on the modulation of mean orbital dynamics, Fig.~\ref{figure_correction_patterns_inclination} presents patterns computed at degree 51 for various inclinations, corresponding to orbits similar to the LRO case. All subplots share identical axis limits, making it evident that the amplitude of the oscillatory components is strongly dependent on the orbital inclination.

By comparing these results with the equipotential structures shown in Fig.~\ref{figure_equipotentials_51}, it becomes clear that an orbit selected for its favorable mean behavior—such as low or frozen mean eccentricity variation—may still exhibit large-amplitude monthly components. These oscillations could significantly degrade the performance in mission profiles sensitive to changes in the eccentricity vector plane, such as stationkeeping.

Furthermore, for frozen or near-frozen lunar orbits, these monthly patterns closely replicate long-term secular behavior and give rise to repeat configurations, as have been previously identified in the literature \citep{RussellLara2007, Lara2011}. For non-frozen orbits, however, these patterns are superimposed on the underlying mean motion drift and must be accounted for accordingly in both trajectory design and maintenance strategies.

\begin{figure}[h]
\centering
\includegraphics[width=\textwidth]{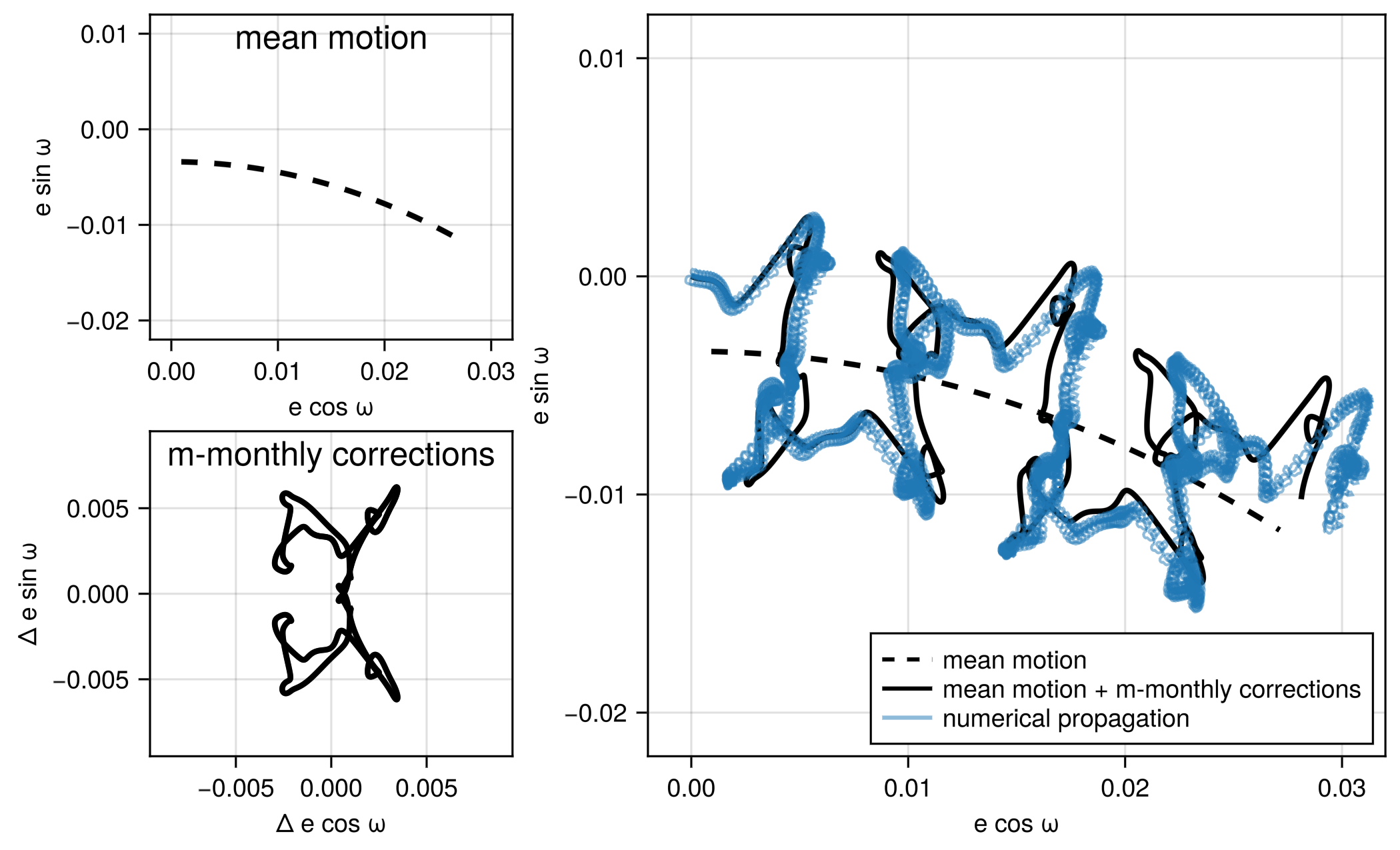}
\caption{Comparing the eccentricity vector propgation of the LRO example using numerical and mean motion + monthly perturbation models.}
\label{figure_lro_translation_theorem}
\end{figure}

An example of this procedure is illustrated in Figure~\ref{figure_lro_translation_theorem} for the LRO case, which does not exhibit frozen mean motion dynamics. The left panels show the decomposition of the eccentricity vector evolution into mean motion and monthly perturbation components. These are then combined and compared against a full numerical propagation in the right panel. The initial conditions and the numerical setup are selected to match those used in the original LRO analysis by \citep{BeckmanLamb2007}, and the resulting dynamical behavior is in close agreement with their findings.

In particular, the superposition of the mean motion and monthly components yields a trajectory that closely reproduces the numerical result. The primary discrepancy arises from the fast oscillations associated with short-period effects, which are removed during the first averaging, but the magnitude of these oscillations is clearly limited.

\section{\label{section_stationkeeping}Stationkeeping using the Translation Theorem}
This section outlines a methodology for automated stationkeeping within \glspl{eLLO}, leveraging the translation theorem applied to numerical propagations. We present the underlying framework, describe the algorithms developed for implementation, and analyze the resulting performance for the three mission examples presented.

\subsection{\label{section_eccentricity}Eccentricity vector control regions}
One of the primary statistics of interest to stationkeeping within \glspl{eLLO} is the eccentricity vector for two primary reasons. Firstly, the stationkeeping specification of spacecraft often requires the maintenance of the orbit between specific altitude bands, which makes the eccentricity vector useful as it encapsulates both the relative magnitudes of periapsis and apoapsis and their corresponding orientations. Secondly, the application of the translation theorem in recent years has made it a useful tool for mission designers. Accordingly, for stationkeeping, we specifically define the orbital regions in which the spacecraft is permitted within the eccentricity vector plane.

These regions within the eccentricity plane must be carefully defined to ensure that stationkeeping constraints are not violated throughout the mission. In this work, \glspl{SDF} are employed to characterize these regions and provide the necessary derivative information for the control optimization. A \gls{SDF} represents the orthogonal distance from a given point to the boundary of a predefined set - which, in this work, the permissible region is defined within the eccentricity vector plane. A positive \gls{SDF} value indicates that the point lies outside the region, whereas a negative value signifies that it remains within. This property allows \gls{SDF} to serve as a basis for developing an algorithm to implement the stationkeeping strategy.

\begin{figure}[h]
\includegraphics[width=\textwidth]{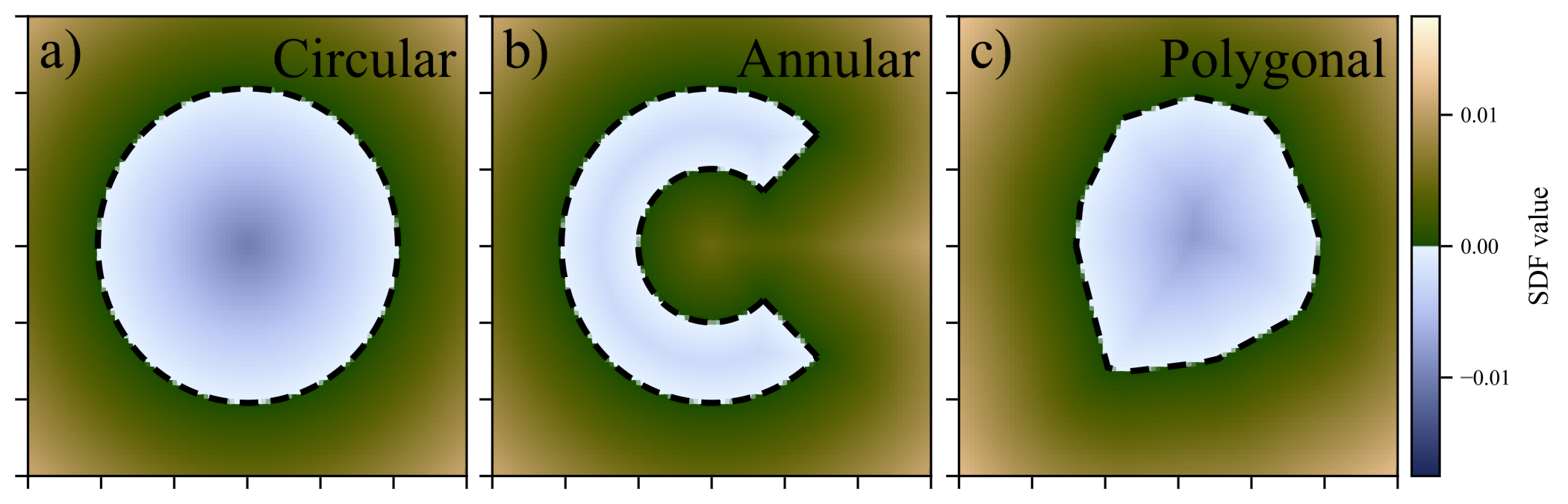}
\caption{\glspl{SDF} for circular, annular/ring and polygonal stationkeeping regions.}
\label{figure_stationkeeping_region_heatmap}
\end{figure}

Fig.~\ref{figure_stationkeeping_region_heatmap} illustrates the three distinct stationkeeping region shapes defined using an \gls{SDF} that are used for the eccentricity vector stationkeeping in this study. Circular stationkeeping regions are applied to cases where a nearly circular orbit is required, which corresponds to a central point on the eccentricity vector plane with zero eccentricity. This is exemplified by the LRO case. Annular (ring-shaped) regions can be employed for more eccentric stationkeeping scenarios, such as in the case of the SER3NE mission. In this mission, the desired orbit maintains a nominal eccentric orbit within a specific range of argument of periapsis. Lastly, polygonal stationkeeping regions can also be used for more generic stationkeeping requirements. These are used particularly for the BINAR mission, where the maximum allowable eccentricity for a specific argument of periapsis varies because of lunar surface collision constraints. The resultant stationkeeping region can be approximated with a polygonal \gls{SDF}.

\begin{figure}[!ht]
\includegraphics[width=\textwidth]{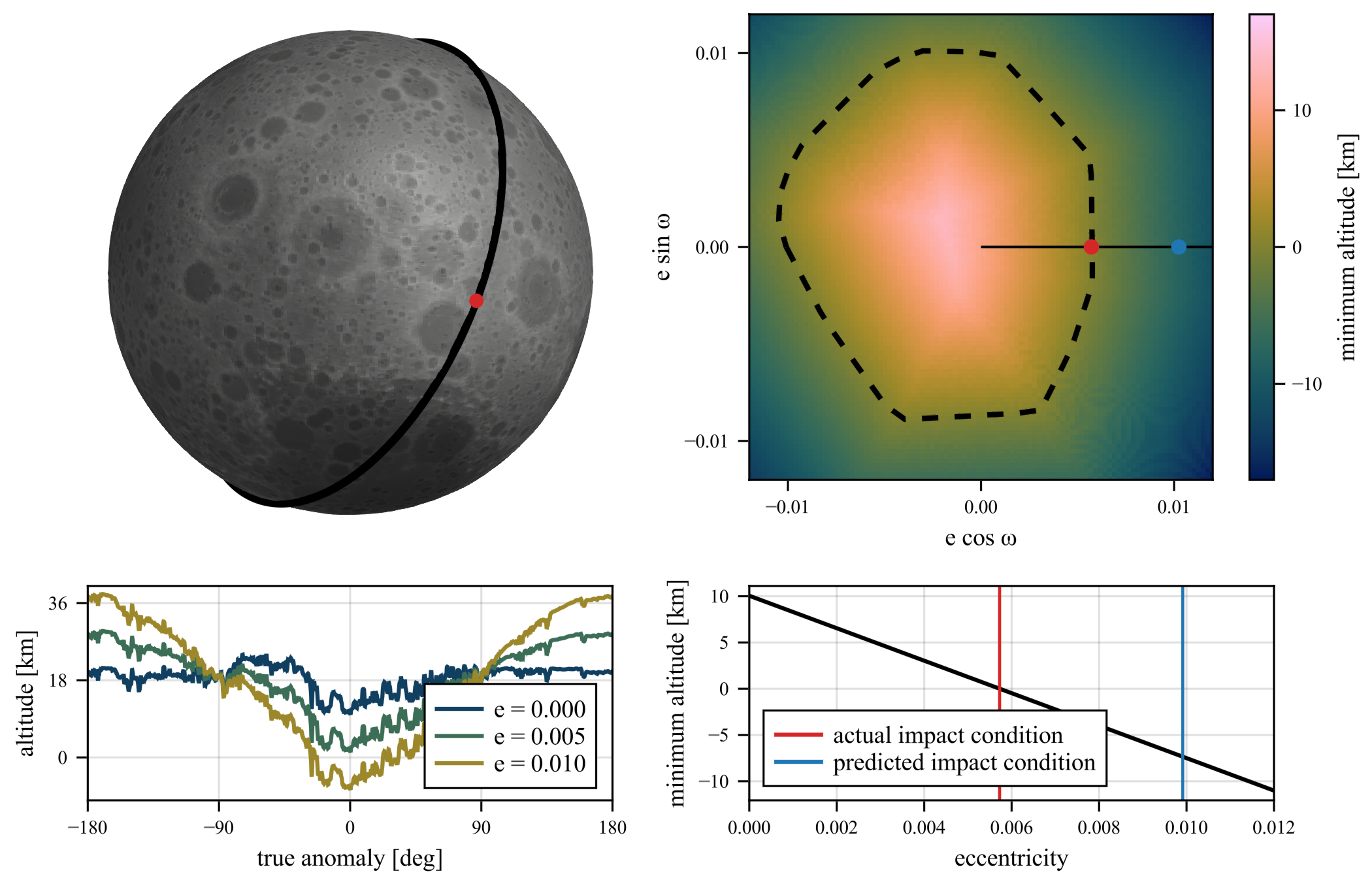}
\caption{Construction of polygonal stationkeeping region for a \gls{eLLO} with $18\si{km}$ altitude and moon-fixed $(\Omega, I)$ of $\ang{210}$ and $\ang{65}$ respectively.}
\label{figure_stationkeeping_region_details}
\end{figure}

\begin{algorithm*}
\caption{\label{algorithm_polygonal_region}Polygonal stationkeeping region generation}
\begin{algorithmic}[1]
\Require $\mathbf{x}, N_f, N_\omega, h_{\text{lim}}$
\State $a,e,I,\Omega,\omega,f \gets \text{CartesianToClassical}(\mathbf{x})$ \Comment{Get current classical elements}
\State $\mathbf{f_N} \gets \text{LinearRange}(0, 2\pi, N_f)$ \Comment{Range of true anomalies to check}
\State $\mathbf{s_h} \gets [\,]$ \Comment{Surface height assuming circular orbit}
\For{$f_{\text{check}} \in \mathbf{f_N}$}
    \State $s_{f} \gets \text{SurfaceHeight}(a, 0, I, \Omega, 0, f_{\text{check}})$ \Comment{Surface height for specific true anomaly}
    \State Append $s_{f}$ to $\mathbf{s_h}$ 
\EndFor
\State $\mathbf{\omega_N} \gets \text{LinearRange}(0, 2\pi, N_\omega)$ \Comment{Range of argument of periapsis to check}
\State $\mathbf{e_{lim}} \gets [\,]$ \Comment{Eccentricity limit for specific $\omega$}
\For{$\omega_{\text{check}} \in \mathbf{\omega_N}$}
    \State Define $J(e) = \text{min}(\mathbf{s_h} - \text{OrbitHeight}(a, e, I, \Omega, \omega_{\text{check}}, \mathbf{f_N})) - h_{\text{lim}}$ \Comment{Rootfinding function}
    \State $e_{\omega} \gets \text{Root}(J(e))$ \Comment{Find eccentricity limit}
    \State Append $e_{\omega}$ to $\mathbf{e_{lim}}$ 
\EndFor
\State $P\gets \text{Polygon}(\mathbf{e_{lim}} \cos \mathbf{\omega_N} , \mathbf{e_{lim}} \sin \mathbf{\omega_N})$ \Comment{Convert to checked values to polygon}
\State \textbf{Return} $\text{SDF}(P)$ \Comment{Return SDF for the polygon}
\end{algorithmic}
\end{algorithm*}

The construction of a polygonal stationkeeping region is illustrated in Fig.~\ref{figure_stationkeeping_region_details}. In the top left, the orientation of the orbital plane with respect to the lunar topography is shown. An orbital plane that intersects the highest topographical features on the lunar far side is deliberately shown. The bottom left shows how the altitude profile of the orbit changes for this orbital plane considering an argument of periapsis of zero; it is clear that the mountainous lunar topology on the lunar far side causes clear jumps in the true altitude. The bottom right pane demonstrates how this causes a large discrepancy between the eccentricity collision predictions based on the circular impact condition with lunar mean radius and the true impact condition. Lastly, the top right pane illustrates the resultant polygonal region by looking at the level set in the eccentricity vector plane where the orbital minimum altitude is $0$ km. 

In order to efficiently generate these polygonal stationkeeping regions, the methodology detailed in Algorithm~\ref{algorithm_polygonal_region} is utilized. This approach systematically evaluates the surface height along the orbit for a range of true anomalies for specific values of the argument of periapsis. This determines a maximum allowable eccentricity for each argument of periapsis, whilst allowing for an altitude margin if required. The resulting region, defined by a sequence of $(\omega, e)$ tuples, can then be transformed into a polygon from which a corresponding \gls{SDF} is generated. Crucially, it is important to consider that this polygonal approximation is only valid for the instantaneous orbital plane of the spacecraft, which is not constant under the lunar rotation and spherical harmonic perturbations. Therefore, the polygonal stationkeeping region is time-dependent and must be recomputed as the trajectories are propagated.

\subsection{\label{section_algorithm}Greedy algorithm to find translation locations}
For any given initial condition that contains a nominal semi-major axis $a$, inclination $I$, and longitude of ascending node $\Omega$, we can additionally set $e=e_\text{ref}$, $ \omega = \omega_{\text{ref}}$, and $f = 0$ in order to propagate the natural evolution of the orbit. These reference values are important because they define the location from which the translation theorem is applied. Therefore, they are selected on the basis of what might be intended by the stationkeeping strategy, for example, from the nominal parameters of the orbit.

Then, by applying the translation theorem in the eccentricity vector plane, it is possible to obtain a good approximation of how the choice of starting location in the eccentricity vector plane would affect the propagation of the eccentricity vector, critically without a full recomputation of the propagation. Therefore, this effect can be utilized in order to find the best starting eccentricity vector offset to maximize the time before the stationkeeping region constraint is violated. 

\begin{figure}[h]
\includegraphics[width=\textwidth]{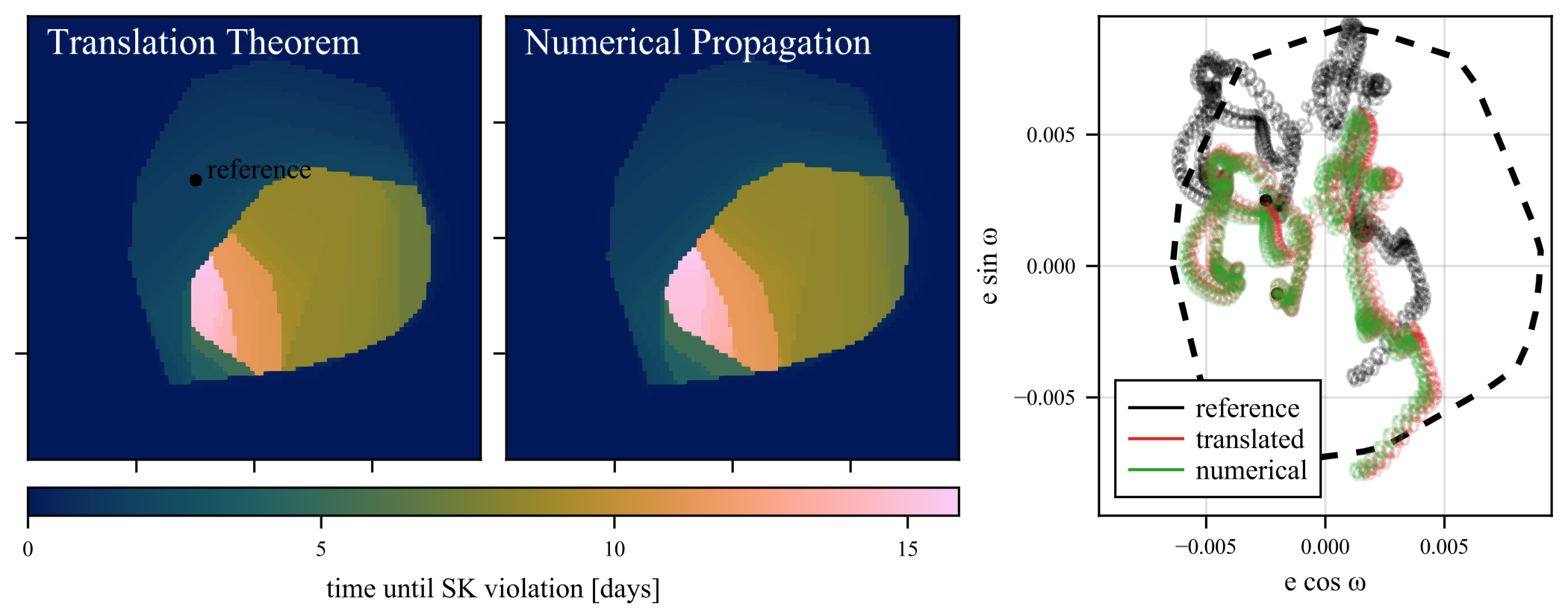}
\caption{Comparing the application of the translation theorem and numerical propagation for the BINAR case.}
\label{figure_stationkeeping_region_comparison}
\end{figure}

Fig.~\ref{figure_stationkeeping_region_comparison} illustrates the validity of the translation theorem for the BINAR case. For visualization purposes, the two heatmaps on the left share identical limits to the plot on the right. The two heatmaps show the predicted duration that each starting location on the eccentricity vector plane remains within the polygonal stationkeeping region, computed using two different methods. In particular, the left heatmap is generated using the translation theorem from the reference point, whereas the heatmap on the right is produced by numerically propagating each starting point individually. The strong agreement between the two heatmaps demonstrates the accuracy of the translation theorem in this scenario.

The rightmost plot provides further insight by visualizing the propagation of selected points on the eccentricity vector plane. The polygonal stationkeeping region is also shown; however, since it is time-dependent, only its initial configuration is displayed. This explains why some apparent violations of the region occur later during the propagation. The black line represents the reference propagation from a circular orbit. The red line shows the propagation of the best-found starting location, assuming the translation theorem applies relative to the reference trajectory. In contrast, the green line shows the same starting point propagated numerically. The close agreement between the red and green trajectories, despite the differing propagation methods, highlights the effectiveness of the translation theorem—particularly when the translations from the reference trajectory are relatively small.

Therefore, it is apparent that the computational advantage of using the translation theorem for this type of grid search is significant. The translation theorem approach requires only a single numerical propagation compared to the need to numerically propagate each grid point. This would make it orders of magnitude more computationally expensive due to the dominance of numerical integration in computational cost, especially when working with high order and degree spherical harmonic models.

\begin{figure}[h]
\centering
    \centering
    \includegraphics[width=3.25in]{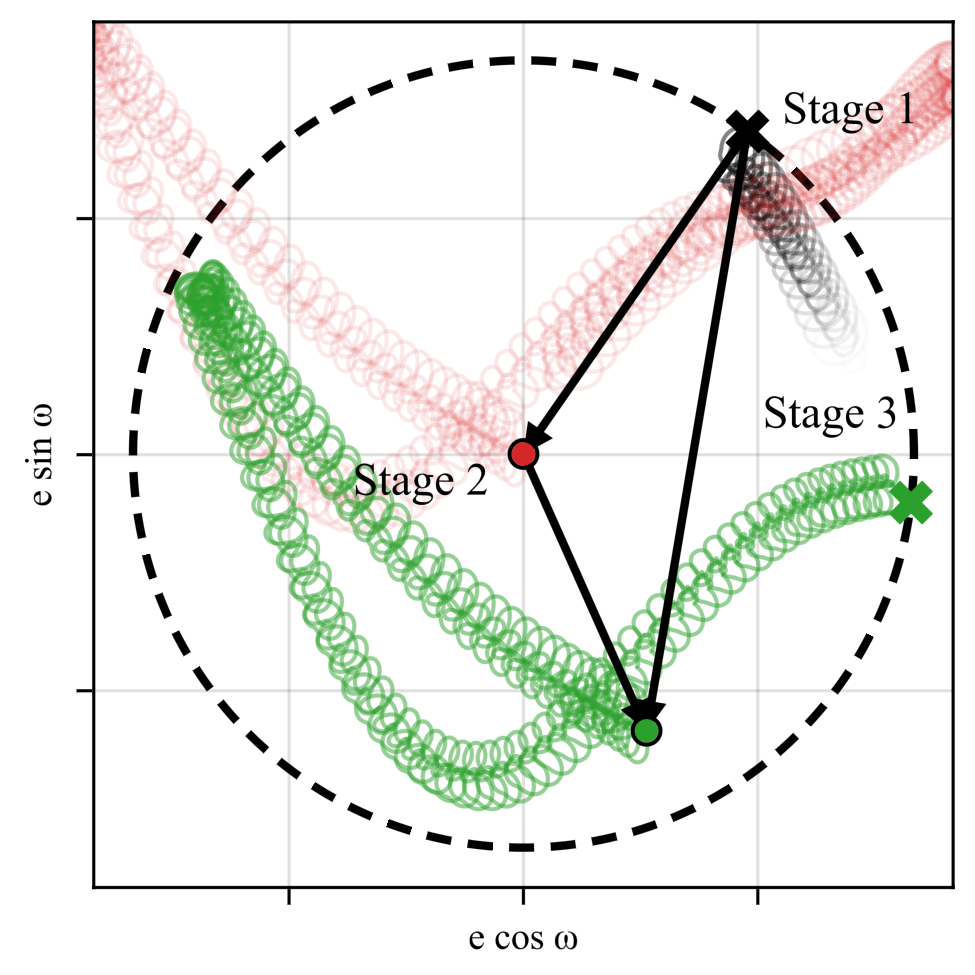}
\caption{\label{figure_ew_explainer}Overview of the stages in finding the best starting location in the eccentricity vector plane.}
\end{figure}

\begin{algorithm*}
\caption{\label{algorithm_ew_optimization}Greedy eccentricity vector stationkeeping}
\begin{algorithmic}[1]

\Require $\mathbf{x}_{\text{start}}, t_\text{start}, t_{\text{end}}, e_\text{ref}, \omega_{\text{ref}}, \Delta t_{\text{step}}, \mathbf{g}_\text{points}$
\State $(\mathcal{C}_\text{ref}, \mathcal{S}_\text{ref}) \gets (e_\text{ref} \cos \omega_{\text{ref}}, e_\text{ref} \sin \omega_{\text{ref}})$ \Comment{Reference eccentricity vector location}
\State $t \gets t_{\text{start}}$ \Comment{Set current time to initial time}
\State $\mathbf{m} \gets []$ \Comment{Storage of maneuver time and location tuples}
\While{$t < t_{\text{end}}$}
    \State $a,e,I,\Omega,\omega,f \gets \text{CartesianToClassical}(\mathbf{x}_\text{start})$ \Comment{Get current classical elements}
    \State $(\mathcal{C}_\text{start}, \mathcal{S}_\text{start}) \gets (e \cos \omega, e \sin \omega)$ \Comment{Starting eccentricity vector location}
    \State $\mathbf{x}_\text{step} \gets \text{ClassicalToCartesian}(a, e_\text{ref}, I, \Omega, \omega_\text{ref}, 0)$ \Comment{Center initial condition on reference location}
    \State $\mathbf{g} \gets \mathbf{g}_\text{points}$ \Comment{Reset possible grid points}
    \State $\mathbf{t}_\text{grid} \gets \text{Fill}(t, \text{Length}(\mathbf{g}))$ \Comment{Time until crash for each grid point}
    \While{$\text{Length}(\mathbf{g}) > 1$ and $t < t_{\text{end}}$} \Comment{Iterate until set is empty / time limit}
        \State $a,e,I,\Omega,\omega,f \gets \text{CartesianToClassical}(\mathbf{x}_\text{step})$
        \State $(\Delta \mathcal{C}_\text{step}, \Delta \mathcal{S}_\text{step}) \gets (e_\text{step} \cos \omega_{\text{step}} - \mathcal{C}_\text{ref}, e_\text{step} \sin \omega_{\text{step}} - \mathcal{S}_\text{ref})$ \Comment{Eccentricity vector reference offset}
        \For{$g \in \mathbf{g}$}
            \State $(\mathcal{C}, \mathcal{S}) \gets g$
            \If {$\text{SDF}(\mathcal{C} + \Delta \mathcal{C}_\text{step}, \mathcal{S} + \Delta \mathcal{S}_\text{step}) \ge 0$} \Comment{Stationkeeping region violation check}
                \State Remove $g$ from $\mathbf{g}$ 
            \Else 
                \State $\mathbf{t}_\text{grid}(g) \gets \mathbf{t}_\text{grid}(g) + \Delta t_\text{step}$ \Comment{Increment time for current grid point}
            \EndIf
        \EndFor
        
        \State $\mathbf{x}_\text{step} \gets \text{Propagate}(\mathbf{x}_\text{step}, \Delta t_{\text{step}})$ \Comment{Integrate until the next time step}
        \State Update stationkeeping region $\text{SDF}$ with new state $\mathbf{x}_\text{step}$ if required
    \EndWhile
    \State $t_\text{next} \gets \text{Maximum}(\mathbf{t}_\text{grid})$
    \State $(\mathcal{C}_\text{target}, \mathcal{S}_\text{target}) \gets \text{Closest}(\mathbf{g}(\mathbf{t}_\text{grid} = t_\text{next}), (\mathcal{C}_\text{start}, \mathcal{S}_\text{start}))$ \Comment{Closest remaining element of $\mathbf{g}$}
    \State $\mathbf{x}_\text{start} \gets \text{ClassicalToCartesian}(a, \text{norm}(\mathcal{C}_\text{target}, \mathcal{S}_\text{target}), I, \Omega, \text{arctan2}(\mathcal{S}_\text{target}, \mathcal{C}_\text{target}), f) $
    \State $\mathbf{x}_\text{start} \gets \text{Propagate}(\mathbf{x}_\text{start}, t_\text{next} - t)$ \Comment{Propagate until next maneuver required}
    \State Append $(t, (\mathcal{C}_\text{target}, \mathcal{S}_\text{target}))$ to $\mathbf{m}$
    \State $t \gets t_\text{next}$
\EndWhile
\State \textbf{Return} m

\end{algorithmic}
\end{algorithm*}

By the repeated application of this process, an eccentricity vector stationkeeping algorithm can be constructed. An overview of this process is illustrated in Fig.~\ref{figure_ew_explainer}, the full details of which are given in Algorithm~\ref{algorithm_ew_optimization}. A simplified summary of the main process of the algorithm is provided below.
\begin{itemize}
    \item Stage 1: A new eccentricity vector translation needs to be found as either (i) it is the initial leg, or (ii) the previous leg has just violated the stationkeeping constraints.
    \item Stage 2: The previous state is translated to a reference location and then this state is propagated until no possible translation would permit the whole trajectory to remain within the stationkeeping constraints.
    \item Stage 3: The translation that is the shortest and then remains the longest time within the stationkeeping region is selected as the next translation target. For the initial leg this location is simply the starting location.
    \item Repeat: Continue propagating after the translation until a stationkeeping violation is found. The process is repeated to form a sequence of eccentricity vector translations until the desired mission duration has been achieved.
\end{itemize}

In order to be computationally efficient, the algorithm uses several discretizations of continuous aspects of the problem. Firstly, the time domain is split such that the stationkeeping constraints are only checked at a certain time frequency, denoted by $\Delta t_\text{step}$. Secondly, the search space within the eccentricity vector plane is also split up so that only translations to a certain discrete number of locations are checked. For the cases in this paper, this discretization was defined to be a rectangular region that encompasses the entire stationkeeping region, which is split by 50 points on each side, for a total of 2500 possible translation locations which are the initial condition $\mathbf{g}_\text{points}$.

When using the eccentricity vector stationkeeping algorithm discussed above, there are two remaining variables for the mission designer to consider when creating a stationkeeping strategy. Firstly, the choice of stationkeeping region is very important. A larger region would, of course, allow for less frequent maneuvers, or perhaps avoid the use of maneuvers altogether. Conversely, a smaller region may cause a large number of maneuvers. Secondly, the choice of initial conditions for the slow variables that do not form the eccentricity vector is critical. These are the semi-major axis $a$, inclination $i$, and longitude of ascending node $\Omega$ (the choice of fast variable $\theta$ does not cause a large difference). An appropriate selection of these initial conditions must be made within the parameters of the mission. 

\subsection{\label{section_conditions}Grid search for best initial conditions}
The low computational cost of the stationkeeping algorithm means that a grid search approach becomes easily accessible in order to evaluate and visualize the ranges of possible initial conditions for the slow variables. Since the semi-major axis $a$ is often fixed by the mission design or configuration, a grid search is constructed on the remaining slow variables, namely the longitude of ascending node $\Omega$ and the inclination $i$. In order to make these results more applicable to a range of different starting times, the grid search particularly scans over the projection of these parameters into the moon-fixed frame. Therefore, despite a set starting time for each mission, the grid search is rather parameterized on the starting configuration of the moon-fixed frame, which can be mapped to any desired time to obtain an approximate stationkeeping profile.

This grid search evaluates the stationkeeping algorithm for each grid point. It is important to note that the assumptions of the stationkeeping algorithm also apply here, the important one being that maneuvers within the eccentricity vector plane are instantaneous. The computational expense of the grid search is largely dependent on the mission duration (directly increasing numerical propagation time) and the number of grid points selected. For example, on an AMD Ryzen 7 5800X3D processor, the grid search for the LRO case with about 5000 grid points took approximately 12 hours using a multi-threaded implementation, which corresponds to a requirement of approximately 8.5 seconds per grid point. 

In the grid search, the two criteria used to quantify the performance of each initial condition are the maneuver count and translation distance, which represent two distinct mission design objectives. These are:
\begin{itemize}
    \item Maneuver count, which is useful for the design of missions where minimizing the number of maneuvers is important, such as in missions where maneuvers effect scientific data collection. Because this is a discrete value, sometimes when calculating the minimum many points on the grid search have the same value. This is resolved by selecting the value with the minimum total translation distance.
    \item Total translation distance, which acts as an approximate metric to quantify stationkeeping $\Delta v$ use. This would be important for fuel-constrained missions, or those which need to spend as long as possible in the desired orbit.
\end{itemize}

\begin{figure}[h!]
\centering
    \centering
    \includegraphics[width=0.9\textwidth]{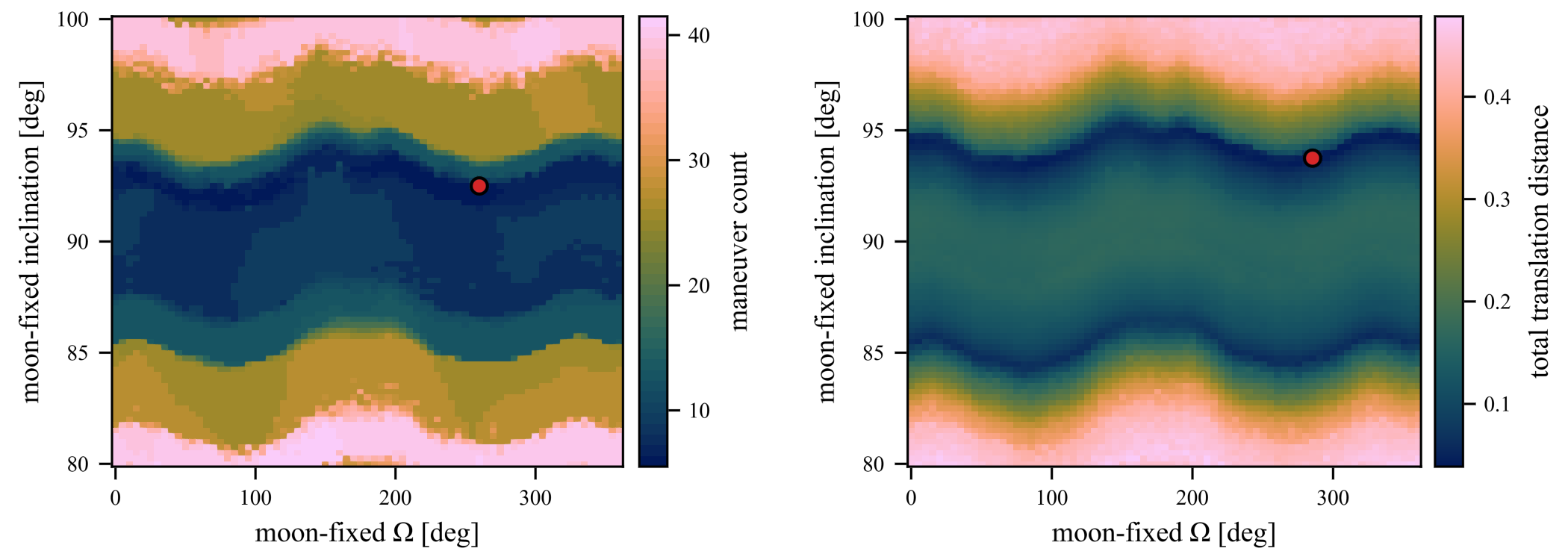}
    \includegraphics[width=0.9\textwidth]{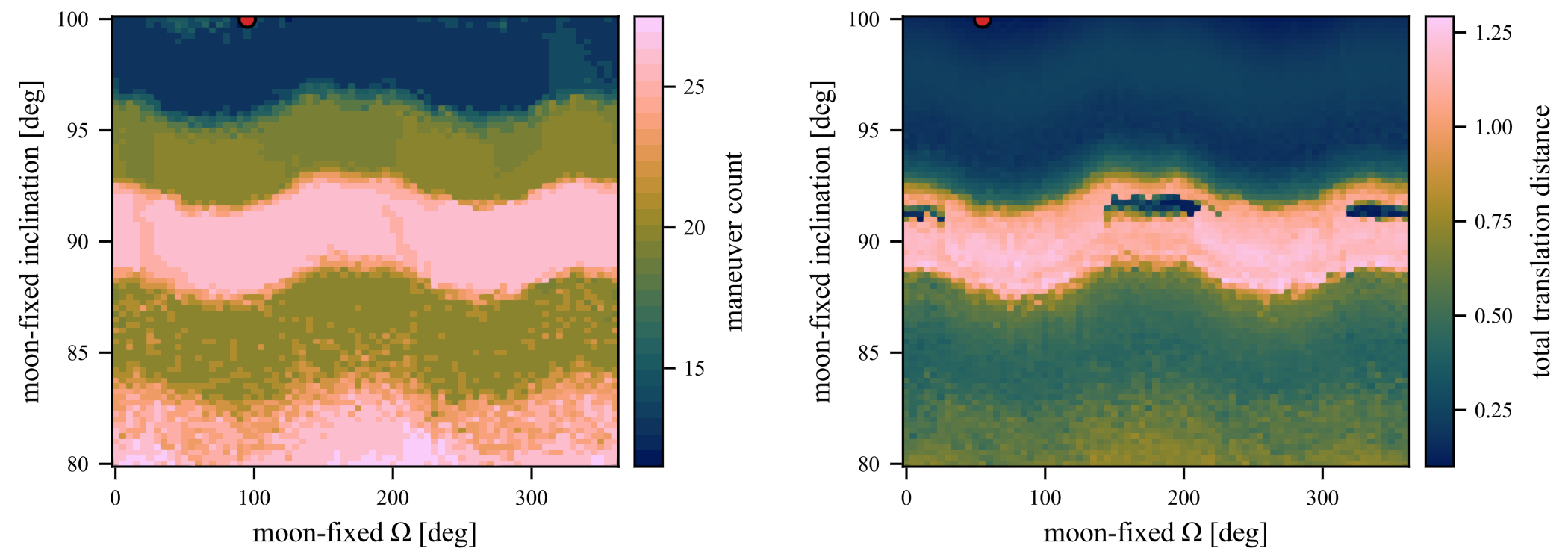}
    \includegraphics[width=0.9\textwidth]{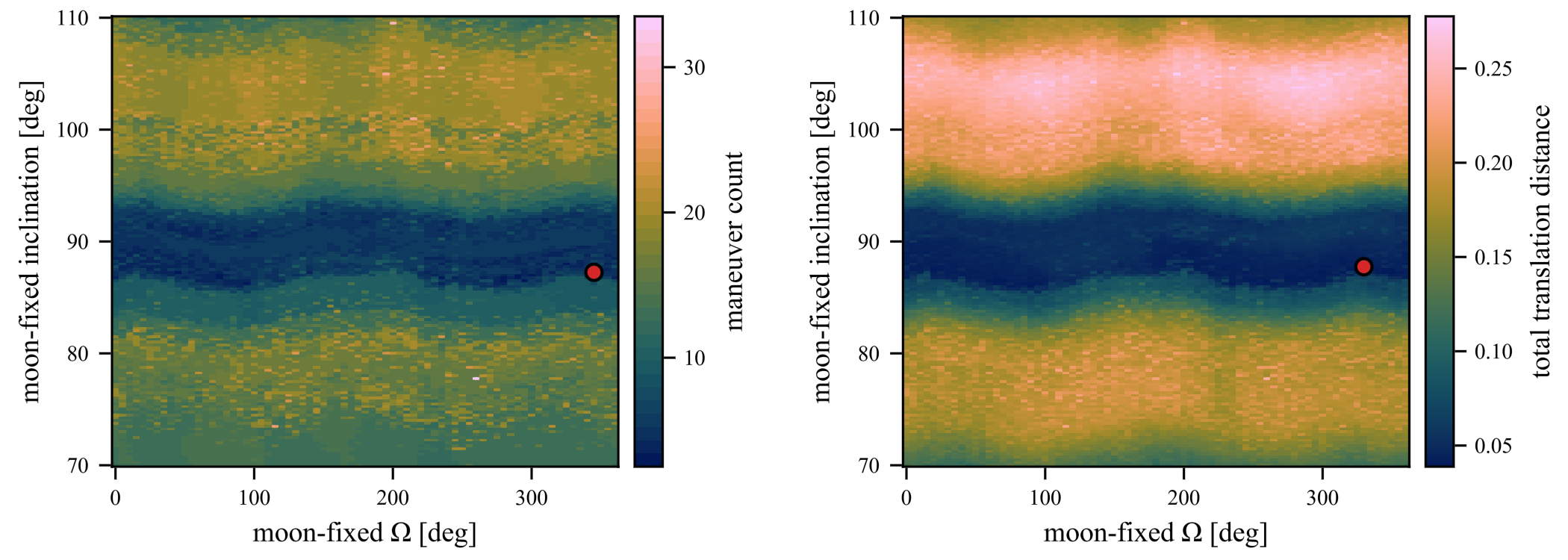}
\caption{\label{figure_grid_search}Grid search results for the LRO (top row), SER3NE (middle row) and BINAR (bottom row) cases.}
\end{figure}

\begin{table}[h]
    \centering
    \begin{tabular}{lccc}
        \toprule
        & \textbf{LRO} & \textbf{SER3NE} & \textbf{BINAR} \\
        \midrule
        \multicolumn{4}{l}{\textit{Minimum maneuver count}} \\
        \quad LAN ($\Omega$) & $\ang{260}$ & $\ang{95}$ & $\ang{345}$ \\
        \quad Inclination & $\ang{92.50}$ & $\ang{100.00}$ & $\ang{87.25}$ \\
        \quad Maneuver count & 6 & 12 & 3 \\
        \quad Translation distance & 0.0877 & 0.1091 & 0.0393 \\
        \multicolumn{4}{l}{\textit{Minimum translation distance}} \\
        \quad LAN ($\Omega$) & $\ang{285}$ & $\ang{55}$ & $\ang{330}$ \\
        \quad Inclination & $\ang{93.75}$ & $\ang{100.00}$ & $\ang{87.75}$ \\
        \quad Maneuver count & 16 & 14 & 4 \\
        \quad Translation distance & 0.0387 & 0.1001 & 0.0388 \\
        \bottomrule
    \end{tabular}
    \caption{The best pairs of moon-fixed LAN and inclination found in the grid search.}
    \label{table_grid_search_results}
\end{table}

The results of the grid search for the three mission cases are illustrated in Fig.~\ref{figure_grid_search}, where the red dots indicate the best initial conditions found for several different criteria. Details on these conditions are provided in Table~\ref{table_grid_search_results}. For each of the mission cases, this grid search has $\ang{0.25}$ spacing on moon-fixed inclination and $\ang{5}$ spacing on the moon-fixed longitude of ascending node. 

In all cases, it is apparent that there is a significant longitude dependence, presenting a clear wavy pattern in the visualization of each grid search. For both the LRO and BINAR cases, both of which have a nominal circular orbit, the best conditions end up placed in a band with inclinations approximately between $\ang{85}$ and $\ang{95}$, which contains several smaller bands that offer the best solutions. Clearly, this analysis would support the conclusions of the designers of these missions that these orbital configurations offer efficient exploration of lunar polar regions. 

In the LRO case, the minimum number of maneuvers found is 6 over the 1-year mission duration with an inclination of $\ang{92.50}$. This is half the number of the initial LRO analysis \citep{BeckmanLamb2007} of 1 maneuver per month for a total of 12, although the latter is for a true polar orbit with strict design requirements. If restricted to true polar orbits, the grid search analysis shows that less than 10 maneuvers per year are possible, with a total translation distance of approximately $0.5$ on the eccentricity vector plane. Comparatively, the initial analysis for LRO demonstrated a translation of approximately $0.08$ per month, which is equivalent to a total of $0.96$.

The BINAR case shows the potential for a mission design that requires between 3 and 4 total maneuvers over the nominal 90-day mission duration, equating to approximately 1 maneuver per month, potentially indicating that the maneuvers are resonant to the monthly pattern. As a result, both the minimum maneuvers and minimum translation distance cases are much more closely aligned than in the LRO example.

The SER3NE case is slightly different and shows that near $\ang{90}$ orbits do not tend to work well; however, interestingly, there seem to be a few very specific regions that, although having large numbers of maneuvers, have low total translation distance. These are nearly the best locations found with the grid search. The small size of these regions stems from the shape of the stationkeeping region, which is quite thin and long, and as such, only orbital configurations with a large quantity of mean and/or monthly motion aligned with the annular shape will have good quality. Due to the limited information available on the SER3NE mission, perhaps these initial conditions are similar to the ones intended, as they agree with the stated mission design of a natural drift in the argument of periapsis \citep{SER3NESELENESEXPLORER}.

\subsection{\label{section_scp_conversion}Maneuver conversion to control profile}
For the grid search procedure, one of the main criteria used to find the best initial conditions is the total translation distance throughout the duration of the mission. This should act as an approximate proxy for the total $\Delta v$ use throughout the mission, and as such, initial conditions with lower total translation distance requirements should generally be cheaper in terms of $\Delta v$. However, for a more accurate assessment of the $\Delta v$ required, a more sophisticated approach using convex optimization techniques is implemented. Applying the general approach of \gls{SCP} \citep{malyutaConvexOptimizationTrajectory2022}, we begin by finding an appropriate linearization of the dynamical system with impulsive maneuvers. Firstly, we define the dynamics for inertial Cartesian state $\mathbf{x}=[\mathbf{r}, \mathbf{v}]$ with a single impulsive maneuver $\mathbf{\Delta v}$ which happens at $t=t_n$,
\begin{align}
    \dot{\mathbf{x}} = f(\mathbf{x}, \mathbf{\Delta v}, t_n) = \left\{\begin{array}{l}
            \dot{\mathbf{r}} = \mathbf{v} \\
            \dot{\mathbf{v}} = g(\mathbf{r}) + \delta(t - t_n) \mathbf{\Delta v} \\
        \end{array}\right.
\end{align}
where $g$ is the gravitational force function calculated from the potential given in Equation~\eqref{equation_gravitational_potential} converted into the inertial frame, and $\delta$ is the Dirac delta function, ensuring the $\mathbf{\Delta v}$ is impulsively applied at $t=t_n$.   

Then, for the linearization procedure, we must split up the trajectory into many initial parts as in a direct method, and each of these parts is referred to as a segment with index $n= 1, 2, ..., N$. The number of segments $N$ is calculated based on an intended timespan between potential impulses. For example, in most of this work, timespans between $10$ minutes and $6$ hours are used depending on the particular situation. 

Subsequently, using this discretization, the linearized dynamic constraints are constructed around a reference trajectory. The \gls{SCP} process requires an appropriate initial guess for the reference trajectory, which should be as close as possible in state to that of the optimal trajectory. For this work, a ballistic reference trajectory is used for most cases. Each segment is assigned a single impulsive maneuver $\mathbf {\Delta v}$. Then, given the reference trajectory ($\mathbf{\bar{x}_n}, \mathbf{\Delta \bar{v}}_n$), the dynamics, and the segmentation $n = 1, 2, ..., N$, a discrete form of the spacecraft dynamics is obtained which can be used as a convex constraint within the \gls{SCP},
\begin{align} \label{equation_dynamics_linear}
    \forall n : \mathbf{x}_{n+1} = \mathbf{A}_n\mathbf{x}_n + \mathbf{B}_n\mathbf{\Delta v}_n + \mathbf{c}_n ,
\end{align}
where,
\begin{align}
    \mathbf{A}_n &= \left. \left[\frac{\partial}{\partial \mathbf{x}} \int_{t_n}^{t_{n+1}}\dot{\mathbf{x}}\,\text{d}t \right]\right|_{(\mathbf{\bar{x}}_n, \mathbf{\Delta \bar{v}}_n)}\\
    \mathbf{B}_n &= \left. \left[\frac{\partial}{\partial \mathbf{\Delta v}} \int_{t_n}^{t_{n+1}} \dot{\mathbf{x}}\,\text{d}t \right]\right|_{(\mathbf{\bar{x}}_n, \mathbf{\Delta \bar{v}}_n)} \\ 
    \mathbf{c}_n &= \mathbf{\bar{x}}_n - \mathbf{A}_n \mathbf{\bar{x}_n} - \mathbf{B}_n \mathbf{\Delta v}_n.
\end{align}

The matrix $\mathbf{A}_n$ is the state transition matrix (STM) representing the changes in the final state $\mathbf{x}_{n+1}$ of each segment with respect to the initial state $\mathbf{x}_{n}$. Correspondingly, $\mathbf{B}_n$ represents the changes in the final state of each segment with respect to the impulsive control $\mathbf{\Delta v}_n$. Note in this case, as the impulsive control is applied at the beginning of each segment, $\mathbf{B}_n$ is identical to the lower half of $\mathbf{A}_n$. 

Rather than using an analytic formulation, the partial derivatives are computed via \gls{AD} which is directly applied to the initial conditions of a numerical integration solver. The VCABM numerical integrator is used from the \texttt{DifferentialEquations.jl} \citep{rackauckasDifferentialEquationsJlPerformant2017} library with absolute tolerance $10^{-10}$ and relative tolerance $10^{-10}$. The \gls{AD} is computed in forward mode through the use of \texttt{ForwardDiff.jl} \citep{revelsForwardModeAutomaticDifferentiation2016}. The use of a multistep numerical integrator was found to improve numerical integration performance (by approximately 2x) in this work due to the high computational cost of evaluating the spherical harmonic gravity model. 

Hard trust region constraints are introduced for the dynamical constraints to ensure that their linearization remains accurate, which are selected to have a constant size that does not change as the \gls{SCP} algorithm progresses. This constraint takes the form
\begin{align} \label{equation_state_trust_regions}
    \forall {n}: -\mathbf{\epsilon}_1 \leq \mathbf{x}_n -\bar{\mathbf{x}}_n \leq \mathbf{\epsilon}_1.
\end{align}
A variety of values for the initial size of the trust regions were tested, but values of $\mathbf{\epsilon}_1$ are selected to be $10 \si{km}$ and $0.01 \si{km/s}$ for position and velocity components respectively. The fixed size of these trust region constraints works well for this problem as they are rarely used because only small deviations from the ballistic trajectory are typically required.

Next, a linearization of the stationkeeping \gls{SDF} function is performed in order to obtain a linear representation of the stationkeeping region constraints. Taking the function $q(\mathbf{x})$ to be the combination of two steps: (i) conversion of the Cartesian state to eccentricity vector and (ii) subsequent evaluation of the \gls{SDF}, this constraint can be expressed as
\begin{align} \label{equation_sdf_constraint}
    \forall n : \mathbf{D}_n\mathbf{x}_n + e_n \leq v_n,
\end{align}
where,
\begin{align}
    \mathbf{D}_n &= \left. \left[\frac{\partial}{\partial \mathbf{x}} q \right]\right|_{(\mathbf{\bar{x}}_n, \mathbf{\Delta \bar{v}}_n)}\\
    e_n &= q(\mathbf{\bar{x}}_n) - \mathbf{D}_n \mathbf{\bar{x}_n}.
\end{align}
As in the dynamical constraint, this linearization is taken around a reference $\mathbf{\bar{x}}_n$ and $\mathbf{\Delta \bar{v}}_n$. The relative simplicity of this process to obtain a convex representation of the stationkeeping region constraints clearly demonstrates an advantage in the use of \gls{SDF} functions for this purpose. The addition of $v_n$ makes this a soft constraint which is penalized in the objective function, of which
\begin{align} \label{equation_virtual_control_positive}
    v_n \ge 0.
\end{align}
This slowly guides the trajectory to meet the stationkeeping region constraints as the \gls{SCP} algorithm progresses. 

Then, the initial state constraint is simply
\begin{align} \label{equation_initial_state}
    \mathbf{x}_{1} = \mathbf{x}_{\text{initial}},
\end{align}
with $\mathbf{x_{\text{initial}}}$ the initial inertial Cartesian state. The target constraint is more complex, as rather than an inertial Cartesian state, a location on the eccentricity vector plane is targeted. Therefore, because the dynamical constraints of the optimization problem are formulated using Cartesian orbital elements, we must also use linearization for the terminal target constraint. Additionally, the semi-major axis is targeted (with overheads) to correct for the drift that may occur if we optimally only target an eccentricity vector location. Then, the terminal target is denoted $\mathbf{y}_\text{target}=[\mathcal{C}, \mathcal{S}, a]$, so the terminal target constraint is
\begin{align} \label{equation_target_ewa}
    \mathbf{F}_N \mathbf{x}_{N} + \mathbf{h}_N = \mathbf{y}_\text{target} + \mathbf{y}_\text{miss},
\end{align}
with,
\begin{align}
    \mathbf{F}_N &= \left. \left[ \frac{\partial}{\partial \mathbf{x}} k \right] \right|_{(\mathbf{\bar{x}}_N)}\\
    \mathbf{h}_N &= k(\bar{\mathbf{x}}_N) - \mathbf{E}_N \mathbf{\bar{x}}_N
\end{align}
where $k(\mathbf{x})$ represents a function which obtains the eccentricity vector and semi-major axis $y=[\mathcal{C}, \mathcal{S}, a]$ from the Cartesian state. The virtual control $\mathbf{y}_\text{miss}$ is added to ensure that the problem is feasible while also allowing for some violation if required and is penalized in the objective function. Then, the absolute value $\mathbf{y}_{\text{miss}}^+$ of each component must be obtained, 
\begin{align} \label{equation_ewa_target_violation}
    \mathbf{y}_{\text{miss}}^+ \ge \mathbf{y}_{\text{miss}} -\mathbf{y}_\text{tol}, &&
    \mathbf{y}_{\text{miss}}^+ \ge -\mathbf{y}_{\text{miss}} -\mathbf{y}_\text{tol},
\end{align}  
where the vector $\mathbf{y}_{\text{tol}}=[10^{-5}, 10^{-5}, 3.0]$ primarily allows for slight variations in each component from that which is targeted. By permitting these variations, the total maneuver cost is kept low as the effects of the fast variations caused by the gravity model are essentially ignored. A larger variation is permitted on the semi-major axis due to its larger relative size (to the eccentricity vector components) and in order to not be largely affected by the fast oscillations.

The main objective will be to minimize the $\Delta v$ use throughout the trajectory. This form of objective necessitates obtaining the 2-norm of the control, which can be achieved through a lossless relaxation via a second-order-cone (SOC) constraint,
\begin{align} \label{equation_soc_dv}
     \Delta v_n \geq ||\Delta\mathbf{v}_n ||_2 \quad (\text{SOC}),
\end{align}
with the 2-norm denoted by the non-bold $\Delta v_n$. Because we are minimizing $\Delta v_n$, this constraint is binding at optimality.

Finally, the objective for the SCP is to minimize the $\Delta v$ use whilst also minimizing the use of virtual controls and soft constraints,
\begin{align} \label{equation_scp_objective}
    J = \sum_{n=1}^N \Delta v_n + 10^2 \sum_{n=1}^N v_n + 10^2 \sum \mathbf{y}_{\text{miss}}^+.
\end{align}
The choice of the value of $10^2$ for the objective penalization tended to work well for our testing. The entire optimization problem is thus
\begin{mini}
    {}{\eqref{equation_scp_objective}}{}{}
    \addConstraint{\eqref{equation_dynamics_linear}}{}{\quad\text{(linearized dynamics)}}
    \addConstraint{\eqref{equation_state_trust_regions}}{}{\quad\text{(state hard trust regions)}}
    \addConstraint{\eqref{equation_sdf_constraint}}{}{\quad\text{(stationkeeping region)}}
    \addConstraint{\eqref{equation_virtual_control_positive}}{}{\quad\text{(positive virtual controls)}}
    \addConstraint{\eqref{equation_initial_state}}{}{\quad\text{(initial state)}}
    \addConstraint{\eqref{equation_target_ewa}}{}{\quad\text{(target derived states)}}
    \addConstraint{\eqref{equation_ewa_target_violation}}{}{\quad\text{(positive target violation)}}
    \addConstraint{\eqref{equation_soc_dv}}{}{\quad\text{(control magnitude)}}
    \label{equation_full_scp_problem}.
\end{mini}
The \gls{SCP} process repeatedly solves \eqref{equation_full_scp_problem} using a convex optimizer and updates the linearized constraints \eqref{equation_dynamics_linear}, \eqref{equation_sdf_constraint} and \eqref{equation_target_ewa} with the optimal solution from the previous iteration. The convergence of the algorithm is determined by the accuracy of the linearization compared to the true one from numerical propagation. In terms of implementation, \texttt{JuMP.jl} \citep{lubinJuMPRecentImprovements2023} is used to create and modify the convex problems, and MOSEK \citep{mosek2025} is used to solve them. 

\subsection{\label{section_scp_application}Application to missions}

\begin{figure}[h]
\centering
    \centering
    \includegraphics[width=\textwidth]{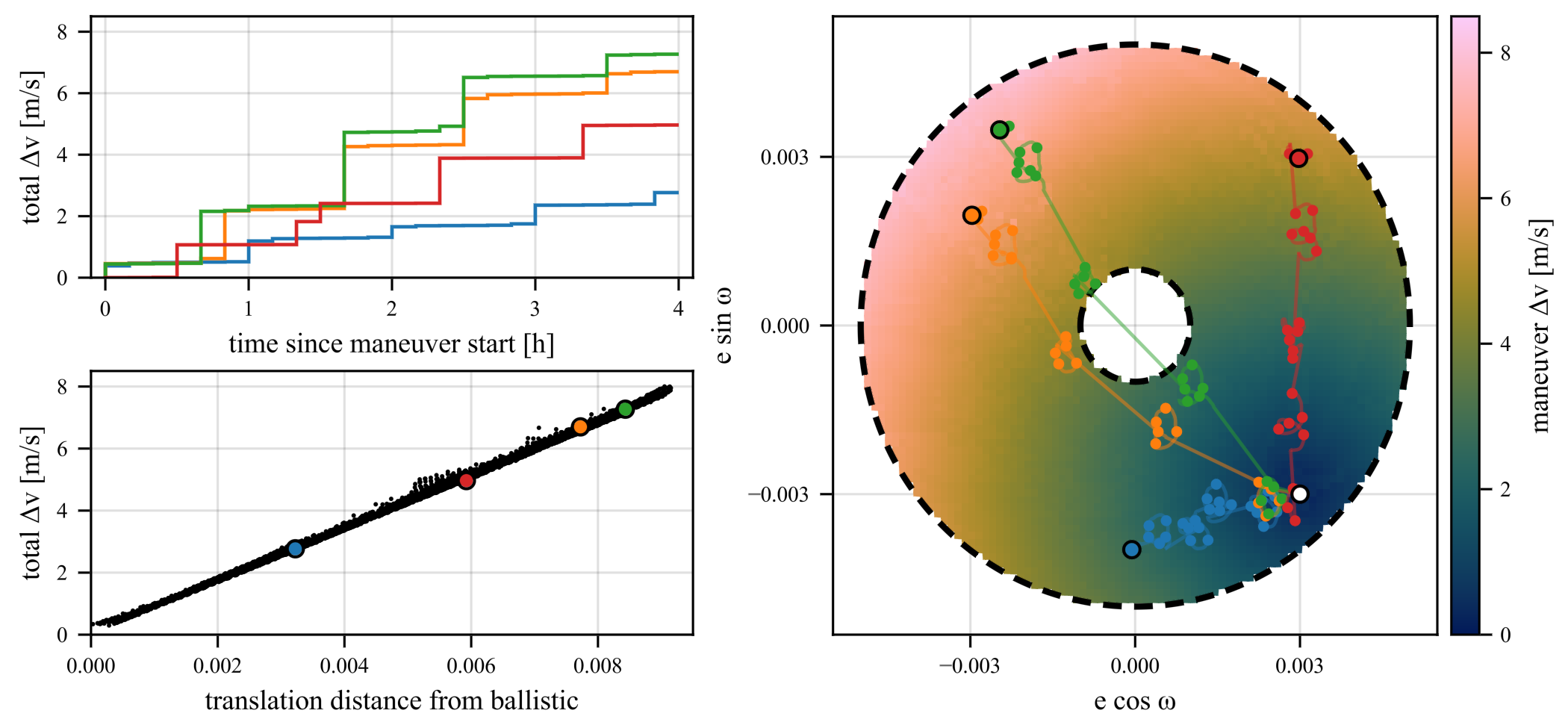}
\caption{\label{figure_translation_cost_heatmap}Computed maneuver sequences and $\Delta v$ cost for a 4-hour translation on the eccentricity vector plane using \gls{SCP}.}
\end{figure}

The \gls{SCP} approach is tested by computing representative maneuvers in the eccentricity vector plane that can be used as replacements for the instantaneous maneuvers that are identified in the grid search stage. This is illustrated in Fig.\ref{figure_translation_cost_heatmap}, where the total $\Delta v$ cost of 4-hour transfers from a particular state on the eccentricity vector plane to other potential states is shown. 
An annular stationkeeping region is used with a 10-minute discretization, which corresponds to a total of 25 nodes in the \gls{SCP}.

The $\Delta v$ cost is near zero for nodes located near the natural propagation, as expected. Furthermore, it is clear that the translation distance serves as an effective proxy for the true $\Delta v$ cost, with the latter exhibiting an approximately linear increase as a function of distance from the endpoint of the natural propagation. The ability of the \gls{SCP} framework in enforcing the \gls{SDF} stationkeeping constraints is demonstrated through the propagation of the green trajectory: although the optimal path seems to pass through the central excluded region, the solution actually avoids violating constraints by executing an instantaneous maneuver directly across it. This behavior persists even under finer temporal discretization, due to the impulsive nature of the control. With non-impulsive control, an appropriately discretized \gls{SCP} formulation would be expected to instead generate trajectories that do not pass through the central forbidden region entirely.

Accordingly, this process of converting instantaneous eccentricity vector translations to inertial control profiles is then used on all initial conditions and their corresponding translation sequences identified in Table~\ref{table_grid_search_results} to generate translation control strategies. In terms of transfer duration, 4-hour transfers are used for the BINAR case, and 8-hour transfers for the LRO and SER3NE cases. It is important to note that since this process considers a fixed time span for the maneuvers, they are started before the predicted translation maneuver such that they both end up at the same eccentricity plane location simultanously. Due to the translation approximation, minor violations of stationkeeping constraints can occur near the end of each phase of the trajectory, so starting the \gls{SCP} maneuvers before the translation is scheduled can actually help improve the adherent constraint in this case. 

For further analysis, these control sequences are then used as an initial guess for \gls{SCP} with discretization throughout the trajectory. This process of generating full-convex control strategies is significantly more computationally intensive primarily due to the increase in the number of discretization nodes and the associated propagation requirements for computing the STM at each iteration of the \gls{SCP} algorithm.

For the LRO and SER3NE mission scenarios, the initial nodes and their corresponding impulsive control inputs derived from the translation-based inertial maneuvers are preserved. Between these, the trajectory is discretized at a 6-hour resolution, resulting in a minimum of 1,460 nodes over the one-year mission duration (four per day). The BINAR scenario is treated similarly, but due to the shorter mission duration, there are a minimum of 360 nodes. The initial nodes, which could have different timespans (e.g., 10 minutes), further augment the problem size. 

Because this formulation allows greater flexibility in determining optimal impulse placement, the resultant cost should be strictly lower than or equal to the translation-based formulation, although at the trade-off of the potentially reduced frequency and length of coasting periods.

In terms of computational cost, the time taken to convert a translation maneuver into a 4-hour maneuver sequence (as in Fig.~\ref{figure_translation_cost_heatmap}) using \gls{SCP} is approximately $5$ seconds. In contrast, for the full-convex \gls{SCP}, the BINAR case requires approximately 30 minutes to converge, while the LRO and SER3NE cases each require approximately 90 minutes, mainly due to the cost of the STM computation and the large-scale nature of the optimization.

\begin{table}[h!]
    \centering
    \begin{tabular}{lccc}
        \toprule
        & \textbf{LRO} & \textbf{SER3NE} & \textbf{BINAR} \\
        \midrule
        \multicolumn{4}{l}{\textit{Minimum maneuver count}} \\
        \quad Translation $\Delta v$ & $76.68$ m/s & $114.86$ m/s & $35.15$ m/s \\
        \quad Translation coast & 98.93\% & 98.39\% & 98.36\% \\
        \quad Full-convex $\Delta v$ & $56.74$ m/s & $81.93$ m/s & $30.67$ m/s \\
        \quad Full-convex coast & 80.83\% & 82.87\% & 91.53\% \\
        \multicolumn{4}{l}{\textit{Minimum translation distance}} \\
        \quad Translation $\Delta v$ & $27.56$ m/s & $77.99$ m/s & $38.38$ m/s \\
        \quad Translation coast & 97.21\% & 98.22\% & 97.81\% \\
        \quad Full-convex $\Delta v$ & $16.81$ m/s & $58.35$ m/s & $31.46$ m/s \\
        \quad Full-convex coast & 91.91\% & 83.67\% & 90.69\% \\
        \bottomrule
    \end{tabular}
    \caption{Results from converting translation sequences to inertial control profiles and the corresponding full-convex SCP.}
    \label{table_final_results}
\end{table}

For each mission scenario, four distinct results are presented. The first two correspond to the translation-based and full-convex refinements of the minimum maneuver count case identified in the grid search. The latter two represent analogous refinements of the minimum translation distance case. A summary of these results is provided in Table~\ref{table_final_results}.

The coast metric is defined as the percentage of mission time during which no maneuvers occur within any sliding window of eight hours. This provides a representative measure of practical spacecraft operation, where long uninterrupted coasting intervals are often preferred to frequent small maneuvers.

In all cases, translation-based solutions yield competitive $\Delta v$ requirements while maintaining excellent coasting characteristics, typically around 98\% of the mission duration. In contrast, full-convex solutions consistently achieve lower $\Delta v$ costs but at the expense of significantly reduced coasting time. For instance, in the case of the BINAR mission with the minimum translation distance, the translation-based solution achieves a total $\Delta v$ of 38.38 m/s, while the full-convex solution achieves a reduction of 18\% to 31.46 m/s. In contrast, the associated coasting durations are 97.81\% and 90.69\%, respectively, indicating that the full-convex solution actively maneuvers for approximately four times longer.

In general, the minimum translation distance configurations tend to perform as well or better than the corresponding minimum maneuver count solutions. An exception seems to be the BINAR scenario, where the minimum translation distance case yields a slightly higher $\Delta v$ than the minimum maneuver count case. However, since both solutions have nearly identical total translation distances (0.0388 vs. 0.0393), and the differences in the resultant full-convex $\Delta v$ are small, this suggests that the discrepancy likely stems from minor variations in orbital geometry or inaccuracies in the translation approximation, rather than a fundamental difference in the methodology between the two solutions.

\begin{figure}[h!]
\centering
    \centering
    \includegraphics[width=\textwidth]{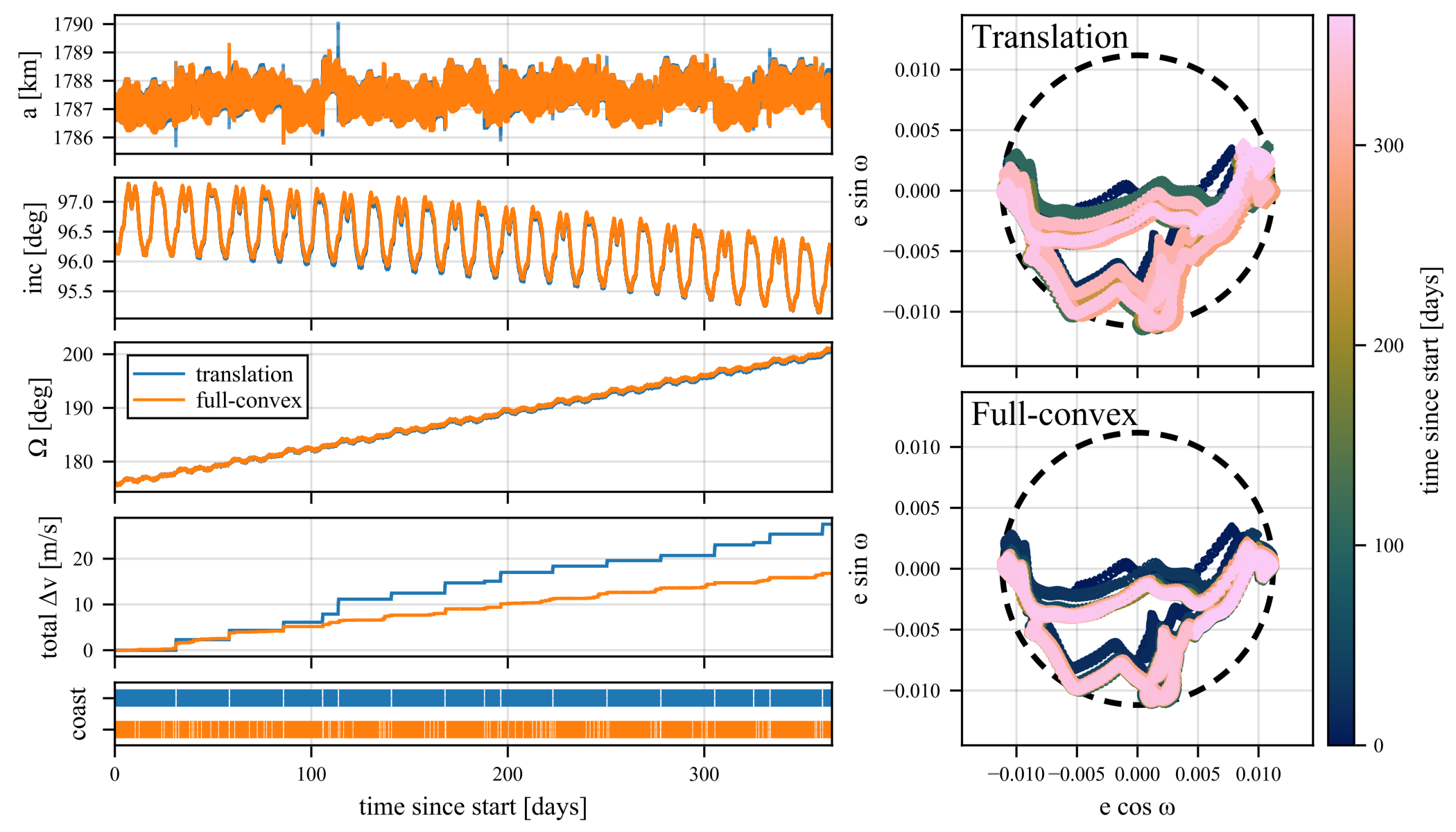}
\caption{\label{figure_translation_convex_lro}Computed trajectories and $\Delta v$ use for the minimum translation distance LRO case.}
\end{figure}

\begin{figure}[h!]
\centering
    \centering
    \includegraphics[width=\textwidth]{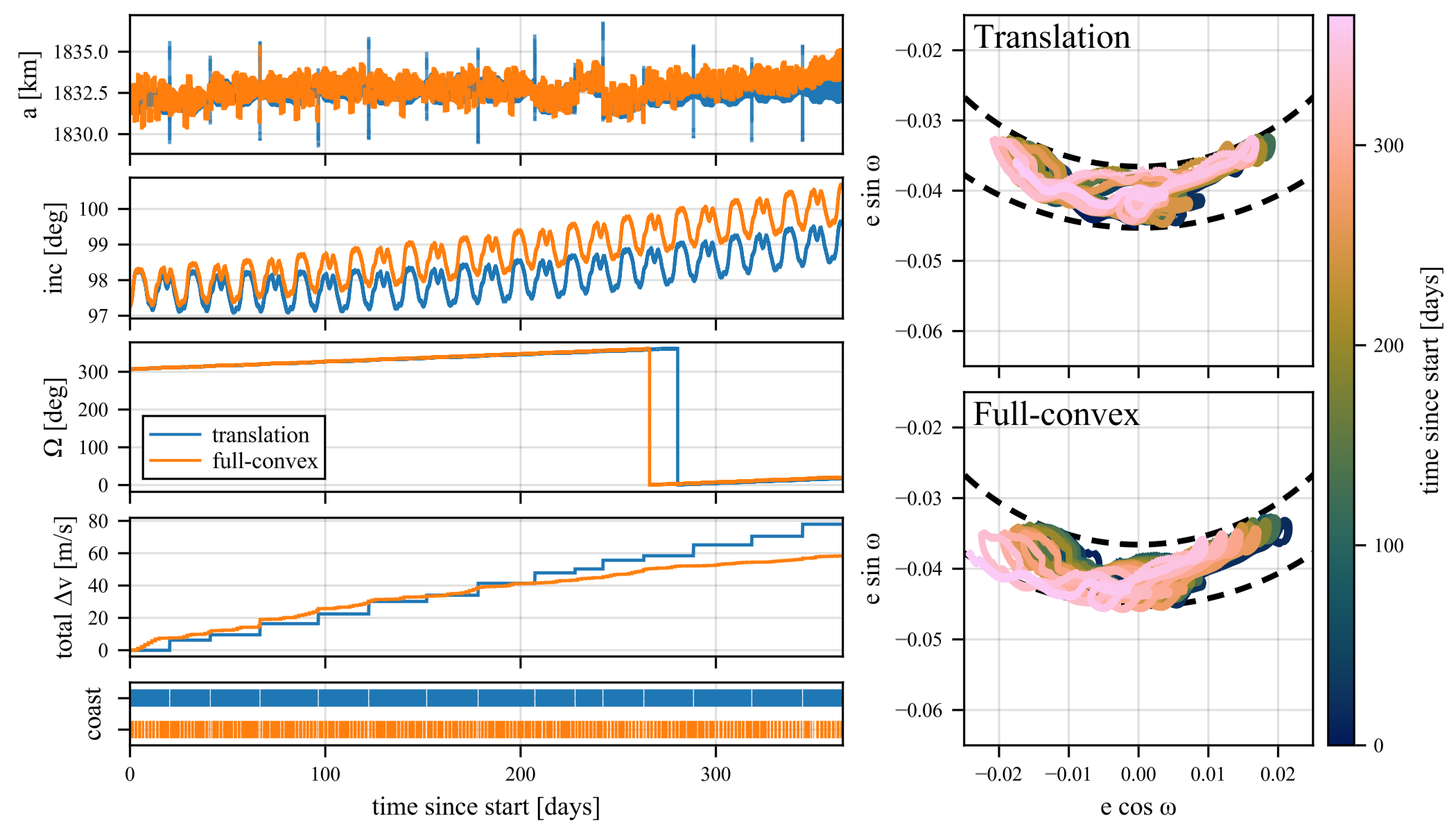}
\caption{\label{figure_translation_convex_ser3ne}Computed trajectories and $\Delta v$ use for the minimum translation distance SER3NE case.}
\end{figure}

\begin{figure}[h!]
\centering
    \centering
    \includegraphics[width=\textwidth]{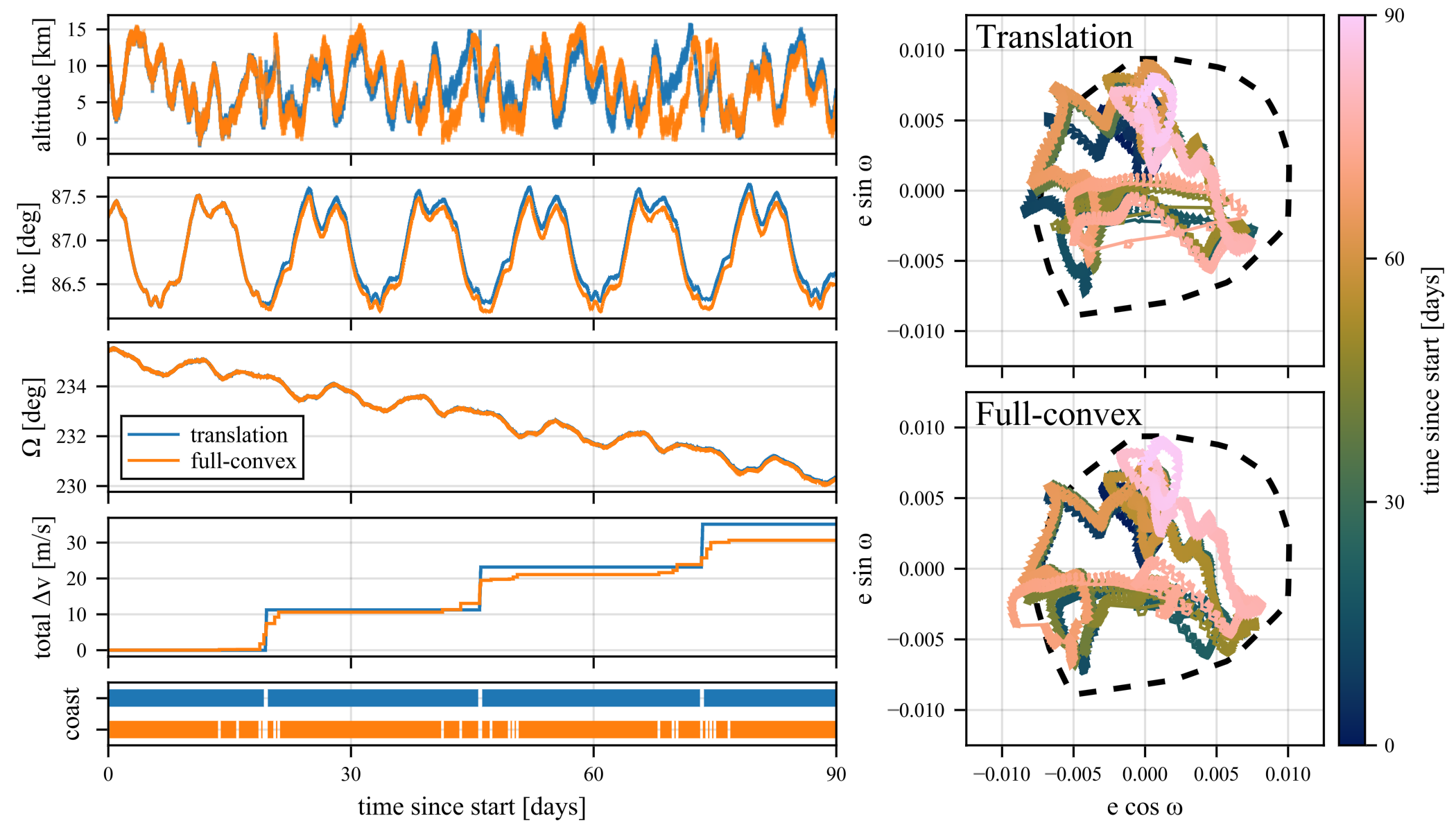}
\caption{\label{figure_translation_convex_binar}Computed trajectories and $\Delta v$ use for the minimum translation distance BINAR case.}
\end{figure}

The results for the minimum translation distance for the LRO and SER3NE cases are are illustrated in Fig~\ref{figure_translation_convex_lro} and Fig.~\ref{figure_translation_convex_ser3ne} respectively. The BINAR case is illustrated with the minimum maneuver count in Fig.~\ref{figure_translation_convex_binar}. In each figure, the left panels present the time evolution of relevant orbital elements in the inertial frame as well as control activity, including thrusting and coasting intervals. The right panels depict the path of the eccentricity vector within the designated control region for both translation-based and full-convex control strategies. This side-by-side comparison enables a direct assessment of the trajectory behavior and stationkeeping adherence for each method.

In the LRO case, it is evident that the algorithm employing the translation theorem has identified a monthly pattern that nearly fits within the stationkeeping region. This explains the relatively low total $\Delta v$ requirements, as only small corrections are needed to maintain the trajectory. The full-convex strategy, which was initialized using the translation-based solution, largely follows a similar path but improves upon it by distributing maneuvers more effectively, causing the monthly movements to almost overlap. The constraints on the drift of the semi-major axis ($\pm3$ km) within the \gls{SCP} are well maintained throughout the mission, while only small drifts are observed in the inertial inclination and longitude of ascending node—parameters which are not explicitly constrained.

The SER3NE case presents particularly interesting behavior due to the annular shape of the control region. It is immediately evident that both the stationkeeping region and the resulting solution behavior differ substantially from those of the LRO case. Here, successful solutions are those in which the mean motion and monthly patterns remain largely aligned within the annular region. Notably, the primary improvements observed in the full-convex case occur during the final portion of the mission, where a controlled drift—both in the orbital elements and on the eccentricity vector plane—is allowed, enabling reductions in the terminal maneuver costs.

In the BINAR case, compared to the other two visualizations, the minimum true altitude over one orbital period of the osculating elements is shown instead of the semi-major axis (consistent with the calculation method used in constructing the polygonal SDF). It is important to emphasize again that the stationkeeping regions are time-dependent; therefore, the regions shown—corresponding to the initial time—do not fully capture the true evolving shape over the mission duration. Both control strategies are clearly very aggressive with respect to the permitted altitudes, but this enables extremely low $\Delta v$ costs for such a low-altitude mission, requiring only approximately monthly maneuvering periods. In some instances, it appears that the trajectories approach very close to or even intersect the lunar surface. Two explanations are proposed for this behavior. For the translation approach, the discrepancy is likely due to inaccuracies introduced by using translated predictions rather than full numerical propagation. For the full-convex approach, it should be noted that the \gls{SCP} node spacing is 6 hours, meaning that potential violations between nodes are not checked for. Reducing the node separation would likely decrease the incidence of such violations, albeit at the cost of significantly increased computational time.

\section{\label{section_conclusions}Conclusions}
\glsresetall
Long-term evolutions of \glspl{eLLO} are affected by significant perturbations that cannot be ignored in orbit design due to their large amplitudes. Some of these perturbations impose peculiar patterns on the eccentricity vector that are barely affected by the mean values of the eccentricity vector itself, which has been referred to as the ``translation theorem'' and have been useful to consider in the design and stationkeeping of \glspl{eLLO}. This analysis sheds light on the roots of the translation theorem, particularly that for small eccentricities, the bulk of the non-zonal perturbations are explained by monthly terms that do not depend on the argument of periapsis and the eccentricity. The combination of the mean motion and these monthly terms can in some instances predict the evolution of the eccentricity vector remarkably well. We believe that with further improvments and a more efficient implementation this could lead to more effective methods of numerically propgating the eccentricity vector.

This work shows that by utilizing the translation theorem an automated algorithm can be constructed for the stationkeeping of \glspl{eLLO} that performs well across a range of possible missions. The computational expense avoided by using the translation theorem permits the large-scale search of possible best initial orbit configurations for stationkeeping whilst keeping the possibility for onboard implementation in spacecraft.

Many avenues remain for further improvements and application of this methodology. In reference to the translation theorem, in some cases the approximation of uniform changes in mean motion across the eccentricity vector plane becomes significant, and we believe that it may be possible to construct higher-order approximations to this to improve accuracy. In terms of the stationkeeping algorithm, the introduction of more operational constraints into this methodology is certainly possible. For example, missed thrust event handling could be resolved by the introduction of a fixed time-to-violation value, where maneuvers are conducted a certain time before violation rather than right as the violation occurs. Other operational constraints, such as fixed maneuver frequencies, single maneuver $\Delta v$ limits, or restricting to periapse and apoapse maneuvers, are all easily applicable to the proposed methodology, but remain to be implemented.

Finally, it is clear that some of the proposed stationkeeping trajectories will not be robust to operational uncertainty, and in the BINAR case this could be fatal to the mission. A future analysis of this methodology under these kinds of uncertainties would be insightful.

\section*{Acknowledgements}

The author(s) would like to thank Robert Howie and Phil Bland from the Space Science and Technology Centre, Curtin University, Perth, Australia, for their early inspiration and collaboration that motivated this work, particularly through their involvement in the BINAR space mission. HH and RA would like to thank Claudio Bombardelli (Universidad Politécnica de Madrid) for the initial brainstorming on this topic during his time as a visiting professor at Auckland.

The contributions by JY, HH, and RA are partially supported by the Royal Society Te Apārangi through the Catalyst: Seeding grant titled ``Advanced Cislunar Space Mission Design''. ML was partially supported by the Spanish State Research Agency and the European Regional Development Fund (Project PID2021-123219OB-I00, AEI/ERDF, EU)

\bibliographystyle{elsarticle-num} 
\bibliography{library.bib}

\end{document}